\documentclass[11pt]{article}
\usepackage{subfigure,amssymb,amsmath,cite,geometry,graphicx}
\pdfoutput=1
\catcode`\@=11

\@addtoreset{equation}{section}
\def\theequation{\arabic{section}.\arabic{equation}}
     
\def\thesection{\arabic{section}}

\def\appendix{\setcounter{section}{0}
        \def\thesection{Appendix.}
        \def\theequation{\Alph{section}.\arabic{equation}}}
\def\section{\@startsection{section}{1}{\z@}{3.5ex plus 1ex minus
   .2ex}{2.3ex plus .2ex}{\large\bf}}

\long\def\@makefntext#1{\parindent 0cm\noindent
\hbox to 1em{\hss$^{\@thefnmark}$}#1}

\newcommand{\captionfonts}{\small}
\makeatletter  % Allow the use of @ in command names
\long\def\@makecaption#1#2{%
  \vskip\abovecaptionskip
  \sbox\@tempboxa{{\captionfonts #1: #2}}%
  \ifdim \wd\@tempboxa >\hsize
    {\captionfonts #1: #2\par}
  \else
    \hbox to\hsize{\hfil\box\@tempboxa\hfil}%
  \fi
  \vskip\belowcaptionskip}
\makeatother   % Cancel the effect of \makeatletter 

\begin{document}

\begin{titlepage}
\vspace{.5in}
\begin{flushright}
October 2014\\  %date
\end{flushright}
\vspace{.5in}
\begin{center}
{\Large\bf
 Black Hole Thermodynamics}\\  %title
\vspace{.4in}
{S.~C{\sc arlip}\footnote{\it email: carlip@physics.ucdavis.edu}\\
       {\small\it Department of Physics}\\
       {\small\it University of California}\\
       {\small\it Davis, CA 95616}\\{\small\it USA}}
\end{center}

\vspace{.5in}
\begin{center}
{\large\bf Abstract}
\end{center}
\begin{center}
\begin{minipage}{4.75in}
{\small
The discovery in the early 1970s that black holes radiate as black bodies has 
radically affected our understanding of general relativity, and offered us some 
early hints about the nature of quantum gravity.  In this chapter I will review 
the discovery of black hole thermodynamics and summarize the many independent 
ways of obtaining the thermodynamic and (perhaps) statistical mechanical properties 
of black holes.  I will then describe some of the remaining puzzles, including the 
nature of the quantum microstates, the problem of universality, and the information 
loss paradox.
}
\end{minipage}
\end{center}
\end{titlepage}
\addtocounter{footnote}{-1}

\section{Introduction}
The surprising discovery that black holes behave as thermodynamic objects has 
radically affected our understanding of general relativity and its relationship to 
quantum field theory.  In the early 1970s, Bekenstein \cite{Bekb,Bekenstein} 
and Hawking \cite{Hawking,Hawkingc} showed that black
holes radiate as black bodies, with characteristic temperatures and entropies 
\begin{align}
kT_{\scriptscriptstyle\mathrm{H}} = \frac{\hbar\kappa}{2\pi} ,\qquad
S_{\scriptscriptstyle\mathrm{BH}}
 = \frac{\ A_{\mathrm{\scriptstyle hor}}}{4\hbar G}  ,
\label{Carlipa1}
\end{align}
where $\kappa$ is the surface gravity and $A_{\mathrm{\scriptstyle hor}}$
is the area of the horizon.  These quantities appear to be inherently quantum 
gravitational, in the sense that they depend on both Planck's
constant $\hbar$ and Newton's constant $G$.   The resulting black body radiation,
Hawking radiation, has not yet been directly observed: the temperature of
an astrophysical black hole is on the order of a microkelvin, far lower than
the cosmic microwave background temperature.  But the Hawking temperature and 
the Bekenstein-Hawking entropy have been derived in so many independent ways,
in different settings and with different assumptions, that it seems extraordinarily
unlikely that they are not real.

In ordinary thermodynamic systems, thermal properties reflect 
the statistical mechanics of underlying microstates.  The temperature of a cup of
tea is a measure of the average energy of its molecules; its entropy is a measure 
of the number of possible microscopic arrangements of those molecules.  It seems 
natural to suppose the same to be true for black holes, in which case the 
temperature and entropy (\ref{Carlipa1}) might tell us something about the 
underlying quantum gravitational states.  This idea has been used as a 
consistency check for a number of proposed models of quantum gravity,
and has suggested important new directions for research.  

In one key aspect, though, black hole entropy is atypical.  For an ordinary 
nongravitational system, entropy is extensive, scaling as volume.  Black 
hole entropy, on the contrary, is ``holographic,'' scaling as area.  If the 
Bekenstein-Hawking entropy really counts black hole microstates, this 
holographic scaling suggests that a black hole has far fewer degrees of 
freedom than we might expect.  But it has also been argued that a black 
hole provides an upper bound on the number of degrees of freedom in 
a given volume: if one tries to pack too many degrees of freedom into a 
region of space,  they may inevitably collapse to form a black hole.  This 
has led to the conjecture that all of Nature may be fundamentally 
holographic, and that our usual counting of local degrees of freedom 
vastly overestimates their number.

Black hole thermodynamics leads to a number of other interesting puzzles as well.  
One is the ``problem of universality.''  If the Bekenstein-Hawing entropy 
$S_{\scriptscriptstyle\mathrm{BH}}$ counts the black hole degrees of 
freedom in an underlying quantum theory of gravity, one might expect that the
result would depend on that theory.  Oddly, though, the expression (\ref{Carlipa1}) 
can be obtained from a number of very different approaches to quantum gravity,
from string theory to loop quantum gravity to induced gravity, in which the
microscopic states appear to be quite different.  This suggests the existence of
an underlying structure shared by all of these approaches,
 but although we have some promising ideas, we do not know what
that structure is.

A second puzzle is the ``information loss paradox.''  Consider a black hole
initially formed by the collapse of matter in a pure quantum state, which then
evaporates completely into Hawking radiation.  If the radiation is genuinely
thermal, this would represent a transition from a pure state to a mixed state, a
process that violates unitarity of evolution and is
forbidden in ordinary quantum mechanics.  If, on the other hand,
the Hawking radiation is secretly a pure state, this would appear to require 
correlations between ``early'' and ``late'' Hawking particles that have never been
in causal contact.  The problem was recently sharpened by Almheiri {\it et al.}
\cite{AMPS}, who argue that one must sacrifice some cherished principle: the
equivalence principle, low energy effective field theory, or the nonexistence of
high-entropy ``remnants'' at the end of black hole evaporation.  These issues
are currently areas of active research, and the matter remains in flux.

\section{Prehistory: black hole mechanics and Wheeler's cup of tea \label{prehist}}

The prehistory of black hole thermodynamics might be traced back to Hawking's
1972 proof that the area of an event horizon can never decrease \cite{Hawkingb}.
This property is reminiscent of the second law of thermodynamics, and the 
correspondence was strengthened with the discovery of analogs of other 
laws of thermodynamics.  This work culminated in the publication
by Bardeen, Carter, and Hawking \cite{BCH}  of the ``four laws of black hole
mechanics'': for a stationary asymptotically flat black hole in four dimensions, 
uniquely characterized by a mass $M$, an angular momentum $J$, and a 
charge $Q$,
%\vspace*{-1.2ex}
\begin{center}
\framebox[5.2in]{\parbox{4.8in}{%
\begin{enumerate}{\addtocounter{enumi}{-1}}
\item The surface gravity $\kappa$ is constant over the event horizon.
\item For two stationary black holes differing only by small  
variations in the parameters $M$, $J$, and $Q$,  
\begin{align}
\delta M 
 = \frac{\kappa}{8\pi G}\delta A_{\mathrm{\scriptstyle hor}} 
+ \Omega_H\delta J + \Phi_H\delta Q ,
\label{Carlipb1}
\end{align}
where $\Omega_H$ is the angular velocity and $\Phi_H$ is the electric 
potential at the horizon.
\item The area of the event horizon of a black hole never decreases,
\begin{align}
\delta A_{\mathrm{\scriptstyle hor}} \ge 0  .
\label{Carlipb2}
\end{align}
\item It is impossible by any procedure to reduce the surface gravity
$\kappa$ to zero in a finite number of steps.
\end{enumerate}
}}
\end{center}

%\vspace*{-1ex}
These laws are numbered in parallel with the four usual laws of thermodynamics,
and they are clearly formally analogous, with $\kappa$ playing the role 
of temperature and $A_{\mathrm{\scriptstyle hor}}$ of entropy.  
As in ordinary thermodynamics, 
there are a number of formulations of the third law, not all strictly 
equivalent; for a summary of the current status, see Wall \cite{Wall}.  
While the four laws were originally formulated for four-dimensional 
``electrovac'' spacetimes, they can be extended to more dimensions, 
more charges and angular momenta, and to other ``black'' objects
such as black strings, rings, and branes.  The first law, in particular, 
holds for arbitrary isolated horizons \cite{Ashisol}, and for much 
more general gravitational actions, for which the entropy can be 
understood as a Noether charge \cite{Wald_Noether}.

But despite the formal analogy with the laws of thermodynamics, 
Bardeen, Carter, and Hawking argued that $\kappa$ cannot be a true
temperature and $A_{\mathrm{\scriptstyle hor}}$ cannot be a
 true entropy.  The defining characteristic
of temperature, after all, is that heat flows from hot to cold.  But if
one places a classical black hole in contact with a heat reservoir
of any temperature, energy will flow into the black hole, but never 
out.  A classical black hole thus has a temperature of absolute zero.

A second thread in the development of black hole thermodynamics began
in 1972, when John Wheeler asked his student, Jacob Bekenstein, what
would happen if he dropped his cup of tea into a passing black hole
\cite{Wheeler}.  The initial state would be a cup of tea (with a nonzero 
entropy) plus a back hole; the final state would be no tea, but a slightly 
larger black hole.  Where did the entropy go?

Bekenstein's ultimate answer was that the black hole must have an entropy
proportional to its area.  Using a series of thought experiments
\cite{Bekb,Bekenstein},  he argued further that the entropy
ought to be of the form 
\begin{align}
S = \eta \frac{A_{\mathrm{\scriptstyle hor}}}{\hbar G} 
\label{Carlipb3}
\end{align}
where $\eta$ is a constant of order one.  For example, a spherical particle 
with a proper radius equal to its Compton wavelength will increase the
entropy (\ref{Carlipb3}) by an amount of order one bit \cite{Bekenstein},
a result that is, remarkably, independent of the spin and charge of the 
black hole.  A small harmonic oscillator dropped into a black hole will 
increase (\ref{Carlipb3}) by an amount at least as great as its entropy
\cite{Bekenstein}, as will a beam of black body radiation \cite{Bekb}.
Bekenstein's ``generalized second law of thermodynamics'' --- the
claim that the total of ordinary entropy of matter plus black hole
entropy $S_{\scriptscriptstyle\mathrm{BH}}$ never decreases --- has
now been tested, and confirmed, over a very wide range of settings
\cite{Wall,Waldb,Wallx}.

\section{ Hawking radiation \label{HR}}

Despite the intriguing parallels between area and entropy, the fact
that a classical black hole had temperature $T=0$ seemed to present an 
insurmountable obstacle to identifying the two.  Zel'dovich had suggested 
in 1970 that quantum effects might be relevant, causing
spinning black holes to radiate in particular modes \cite{Zeldovich}, 
but the arguments were not worked out in detail.  In 1974, however, 
Hawking \cite{Hawking,Hawkingc} used newly developed techniques
 for treating quantum field  theory in curved spacetime \cite{Parker} 
to show that \emph{all} black holes radiate, with
a black body temperature $T_{\scriptscriptstyle\mathrm{H}}$.  With this
identification of temperature, the first law of black hole mechanics
(\ref{Carlipb1}) determines the entropy, and
Bekenstein's expression (\ref{Carlipb3}) is confirmed, with
$\eta=1/4$.\footnote{Strictly speaking, the first law determines 
black hole entropy only up to an additive constant. But Pretoruis
{\it et al.}\ have designed an (idealized) process that \emph{reversibly} 
forms a black hole with entropy $S_{\scriptscriptstyle\mathrm{BH}}$, 
strongly hinting that no additive constant appears \cite{Pretorius}.}

There are several ways to describe Hawking's results.  Perhaps the most
intuitive is to say that quantum mechanics allows particles to tunnel out
through the event horizon; but while Hawking himself used this as
a heuristic picture \cite{Hawkingc}, the full mathematical
description of such a tunneling process was not worked out until much 
later \cite{Parikh}.  Hawking's calculation was based, rather, on a key
feature of quantum field theory in curved spacetime: the fact that the vacuum
is not unique, but depends on a choice of time.  In particular, Hawking
showed that the Minkowski vacuum of a past observer watching the 
collapse of a star differed from the vacuum of a future observer looking
at the resulting black hole.

\subsection{Quantum field theory in curved spacetime}

To be more explicit,  consider a free massless
scalar field $\varphi$.  In  Minkowski space, the standard approach
to quantization \cite{BirrellDavies} is to first expand $\varphi$ in its 
Fourier modes,
\begin{align}
\varphi = \sum_{\bf k} \left(a_{\bf k}f_{\bf k}(t,{\bf x}) 
  + a_{\bf k}^\dagger f_{\bf k}^*(t,{\bf x})\right) 
  \quad \mathrm{with}
  \ f_{\bf k} = c_{\bf k}e^{i{\bf k}\cdot{\bf x} - i\omega_{\bf k}t},\ 
  \omega_{\bf k} =  |{\bf k}|   .
\label{Carlipc1}
\end{align}
The $f_{\bf k}$ are positive frequency, positive energy modes,
which may be completely characterized by the conditions
\begin{align}
\Box  f_{\bf k}(t,{\bf x}) = 0, \quad 
 \partial_t f_{\bf k}(t,{\bf x}) 
   = -i\omega_{\bf k}f_{\bf k}(t,{\bf x}) \qquad\hbox{(with $\omega_{\bf k}>0$)} .
\label{Carlipc2}
\end{align}
Similarly, the $f_{\bf k}^*$  are negative frequency, negative energy modes.
The theory has a natural scalar product,
\begin{align}
(\varphi_1,\varphi_2) = -i\int_\Sigma n^a(\varphi_1
   \overset{\leftrightarrow}{\partial}_a\varphi_2^*)\sqrt{g_{\scriptscriptstyle\Sigma}}
   \,d^3x,
\label{Carlipc2a}
\end{align}
where $\Sigma$ is an arbitrary Cauchy surface  and $n^a$ is its unit normal.    
The modes (\ref{Carlipc2}) are then orthonormal, provided that we choose the
normalization constants
\begin{align}
c_{\bf k} = \frac{1}{\sqrt{(2\pi)^3 2\omega}}  .
\label{Carlipc2b}
\end{align}

To quantize the theory, we now interpret the coefficients of the $f_{\bf k}$ as 
annihilation operators, and the coefficients of the $f_{\bf k}^*$ as creation 
operators.  These obey the standard commutation relations, and can be
used to construct the usual Fock space of free particle states.  In particular,
the vacuum is the state annihilated by the $a_{\bf k}$,
\begin{align}
a_{\bf k} |0\rangle = 0  .
\label{Carlipc3}
\end{align}

In a more general spacetime, 
or even a noninertial frame in Minkowski space, the standard 
Fourier modes no longer exist.  Suppose, though, that we can still choose
a  preferred time coordinate $t$.  We can then find a complete set 
of orthonormal modes satisfying (\ref{Carlipc2}),  perform an expansion 
of the form (\ref{Carlipc1}) with respect to these modes, and once again 
define creation and annihilation operators and a vacuum state.  

What is new, though, is that there may now be more than one preferred time.
Given two choices, say $t$ and $t'$, two expansions exist:
\begin{align}
\varphi = \sum_i \left(a_if_i + a_i^\dagger f_i^*\right)  
  = \sum_i\left(a'_if'_i + {a'_i}^\dagger {f'_i}^*\right)   .
\label{Carlipc4}
\end{align}
Furthermore, since the $\{f_i,f_i^*\}$ are a complete set of functions, we can
write
\begin{align}
f'_j 
  = \sum_i \left(\alpha_{ji}f_i + \beta_{ji}f_i^*\right) 
\label{Carlipc5}
\end{align}
for some coefficients $\alpha_{ji}$ and $\beta_{ji}$.
This relation is known as a Bogoliubov transformation, and 
the coefficients $\alpha_{ji}$ and $\beta_{ji}$ are Bogoliubov 
coefficients \cite{Bogoliubov}.   The coefficients may be
read off from (\ref{Carlipc5}) by using the inner product (\ref{Carlipc2a});
in particular,
\begin{align}
\beta_{ji} = -(f^*_i,f'_j)  .
\label{Carlipc5a}
\end{align}

We now have two vacua, a state $|0\rangle$ annihilated by the 
$a_i$ and a state $|0'\rangle$ annihilated by the $a'_i$, and
two number operators, $N_i=a_i^\dagger a_i$ 
and $N'_i={a'_i}^\dagger a'_i$.  Using (\ref{Carlipc5}) and the 
orthonormality of the mode functions, it may be easily checked that
\begin{align}
\langle 0'| N_i |0'\rangle = \sum_j |\beta_{ji}|^2 .
\label{Carlipc6}
\end{align}
Thus if the coefficients $\beta_{ji}$ are not all zero, the 
``primed'' vacuum will have a nonvanishing ``unprimed'' particle 
content.

\subsection{Hawking's calculation \label{Hawka}}

In his seminal work on black hole radiation and evaporation 
\cite{Hawking,Hawkingc}, Hawking computed the Bogoliubov 
coefficients between an initial vacuum outside a collapsing star 
and a final vacuum after the formation of a black hole.  He showed 
that a ``primed'' observer in the distant future would see a thermal 
distribution of particles at the Hawking temperature (\ref{Carlipa1}).  
While Hawking's full calculation is too technical to include
in this review --- see, for example, Traschen's very nice paper
\cite{Traschen} for details --- it is possible to sketch out the basic
argument.
\begin{figure}[ht]
\centerline{
\subfigure[Carter-Penrose diagram for a star collapsing to form a black hole]
{\includegraphics[width=1.75in]{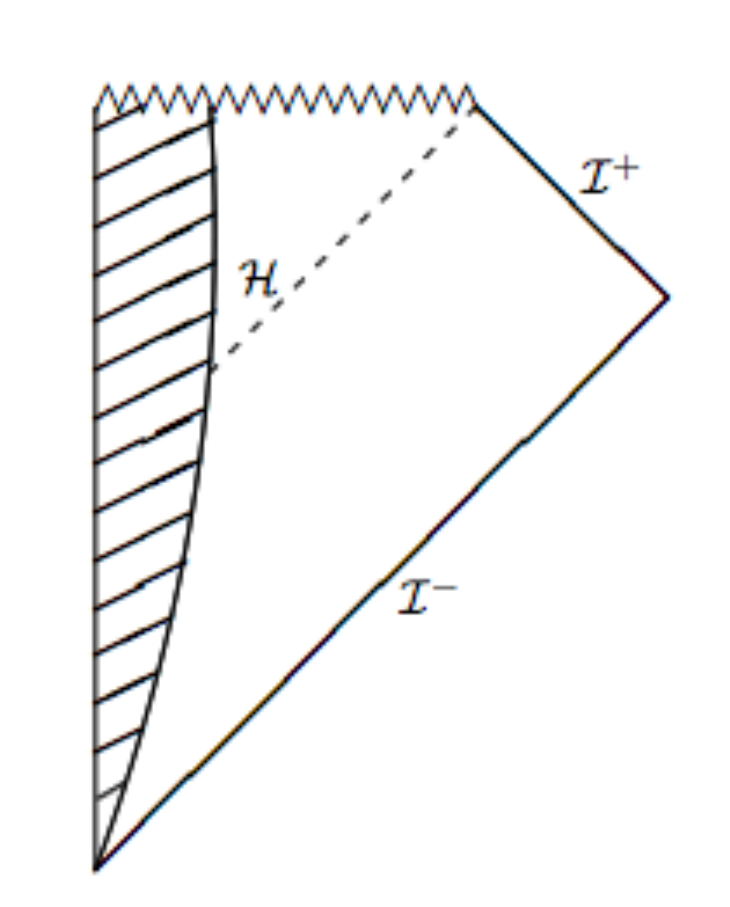}\label{Carlipfig0a}}
\hspace*{30pt}
\subfigure[Coordinates, ray-tracing, and mode functions]
{\includegraphics[width=1.6in]{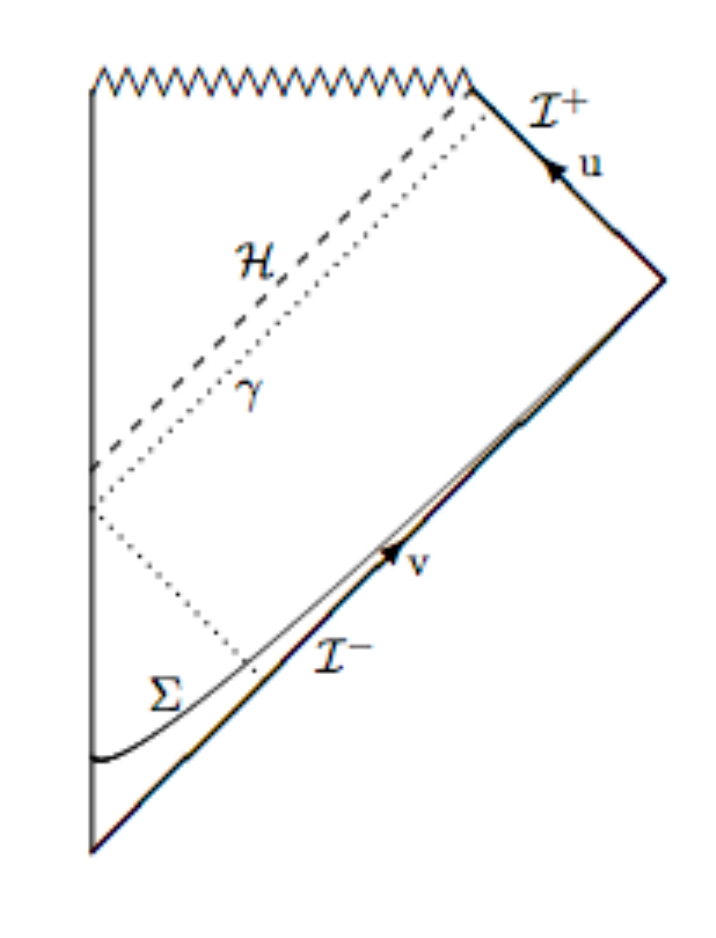}\label{Carlipfig0b}}
}
\caption{The set-up for Hawking's calculation.} \label{Carlipfig0}
\end{figure}

Hawking's starting point is the collapse of a star to form a black hole,
as shown in the Carter-Penrose diagram of figure \ref{Carlipfig0a}. The
shaded area is the collapsing star; the horizon $\mathcal{H}$ forms at
some time after the collapse has begun.  
In the distant past, the spacetime far from the star is nearly flat, and we
can define a standard Minkowski vacuum at past null infinity $\mathcal{I}^-$.
In the distant future, the black hole has settled down to a (nearly) stationary
configuration, and we can define a vacuum at future null infinity $\mathcal{I}^+$.
From the previous discussion, we need to find the orthonormal modes 
at $\mathcal{I}^\pm$, propagate them to a common Cauchy surface --- 
$\Sigma$ in figure\ref{Carlipfig0b} --- and take the inner product (\ref{Carlipc5a}) 
to determine the Bogoliubov coefficients.

For the first step, let us choose null coordinates $u$ (``retarded time'')
and $v$ (``advanced time'') in the exterior of the black hole, chosen so
that $u$ is an affine parameter along $\mathcal{I}^+$ and $v$ is an
affine parameter along $\mathcal{I}^-$, as shown in figure \ref{Carlipfig0b}.
It is easy to check that the affine condition implies that near
$\mathcal{I}^\pm$, the metric takes the form $ds^2 = dudv + \hbox{angular
terms}$.  As a result, the Klein-Gordon equation greatly simplifies, and the
asymptotic modes become
\begin{alignat}{3}
&f_{\omega \ell m} \sim \frac{1}{r}Y_{\ell m}(\theta,\phi)e^{-i\omega v} 
  \qquad &&\hbox{near $\mathcal{I}^-$} ,\nonumber\\
&f'_{\omega \ell m} \sim \frac{1}{r}Y_{\ell m}(\theta,\phi)e^{-i\omega u} 
  \qquad &&\hbox{near $\mathcal{I}^+$}   .
\label{Carlipc7}
\end{alignat}

To calculate the inner product (\ref{Carlipc5a}), Hawking evolved 
the mode $f'_{\omega \ell m}$ from $\mathcal{I}^+$ backwards to 
$\mathcal{I}^-$.  In principle, this requires solving the Klein-Gordon equation 
with boundary conditions at $\mathcal{I}^+$.   For the relevant frequencies, 
though, the geometric optics approximation, in which massless particles move 
along null geodesics, is good enough.   The dotted line labeled $\gamma$
in figure \ref{Carlipfig0b} shows a null geodesic propagating backwards 
from $\mathcal{I}^+$, ``reflecting'' off $r=0$, and continuing to 
$\mathcal{I}^-$.  (One may visualize this as an ingoing spherical 
wave shrinking to a point at the origin, passing through itself, and expanding 
outward.)  This geodesic maps a point $u$ on $\mathcal{I}^+$ to a 
point $p(u)$ on $\mathcal{I}^-$, and the mode functions $f'_{\omega \ell m}$
in turn map to functions of the form
\begin{align}
\frac{1}{r}Y_{\ell m}(\theta,\phi)e^{-i\omega p^{-1}(v)}   .
\label{Carlipc8}
\end{align}

The meat of Hawking's calculation is the determination of this function
$p(u)$, which he finds by tracking the position of the geodesic
$\gamma$ relative to the horizon $\mathcal{H}$.   The result is that
\begin{align}
p(u) \sim A - Be^{-\kappa u}  ,
\label{Carlipc9}
\end{align}
where $\kappa$ is the surface gravity and $A$ and $B$ are constants.  The 
exponential dependence on $u$ reflects the extreme blue shift of waves
passing near the horizon: two successive crests at $\mathcal{I}^+$,
labeled by $u$ and $u+\delta u$, will be mapped to a separation 
$p'(u)\delta u \sim \kappa Be^{-\kappa u}\delta u$ at $\mathcal{I}^-$.
Viewed in the opposite direction, the function $p(u)$ describes the
``peeling'' of null geodesics away from the horizon \cite{BLSV}.

The computation of the Bogoliubov coefficients is now straightforward.
The inner product (\ref{Carlipc5a}) on a Cauchy surface $\Sigma$ near 
$\mathcal{I}^-$ now involves integrals of the form
\begin{align*}
\int dv\,e^{i\omega' v} e^{-i\frac{\omega}{\kappa}\ln v}     .
\end{align*}
These yield gamma functions of complex arguments, whose absolute 
squares give the exponential behavior of a thermal distribution
with the Hawking temperature.

I have left out many steps, of course.  In
particular, future null infinity $\mathcal{I}^+$ is not a Cauchy surface,
so the modes $f'_{\omega \ell m}$ in (\ref{Carlipc7}) are not a complete
set; they must be supplemented by modes on the horizon $\mathcal{H}$.
These account for backscattering into the
black hole.  This scattering is energy-dependent, and distorts the
black body spectrum, yielding instead a ``grey body spectrum''
\begin{align}
\langle N\rangle = \frac{\Gamma_\omega}{e^{\frac{2\pi\omega}{\kappa}}-1}
\label{Carlipc10}
\end{align}
where the reflection coefficient $\Gamma_\omega$ can be calculated 
explicitly \cite{Page,Page2}.  Generalizations to particles other than massless
scalars are straightforward, and give a consistent thermodynamic picture.

\section{Back-of-the-envelope estimates}

The description I have given of Hawking's derivation of black hole radiation
is not, for most people, terribly intuitive.  It is useful to understand
some back-of-the-envelope estimates, which can give a better feel
for the physics.  The following examples are taken from
a set of my earlier lectures \cite{Carlip}, slightly updated.

\subsection{Entropy \label{Enta}}
We start with an estimate of black hole entropy, using the generalized
second law \cite{Kiefer}.  Consider a cubic box of gas with volume
$\ell^3$, mass $m$, and temperature $T$, dropped into a Schwarzschild 
black hole of mass $M$ (and therefore having a horizon area of
$A = 16\pi G^2M^2$).  Let us assume that the box is at least as large 
as the thermal wavelength of the gas, $\ell\sim \hbar/T$.  Then the 
descent of the gas into the black hole will lead to a decrease of entropy
\begin{align}
\Delta S_{\mathrm{\scriptstyle gas}} \sim -\frac{m}{T} \sim -\frac{m\ell}{\hbar} .
\label{Carlipd1}
\end{align}
\begin{figure}
\centerline{\includegraphics[width=4in]{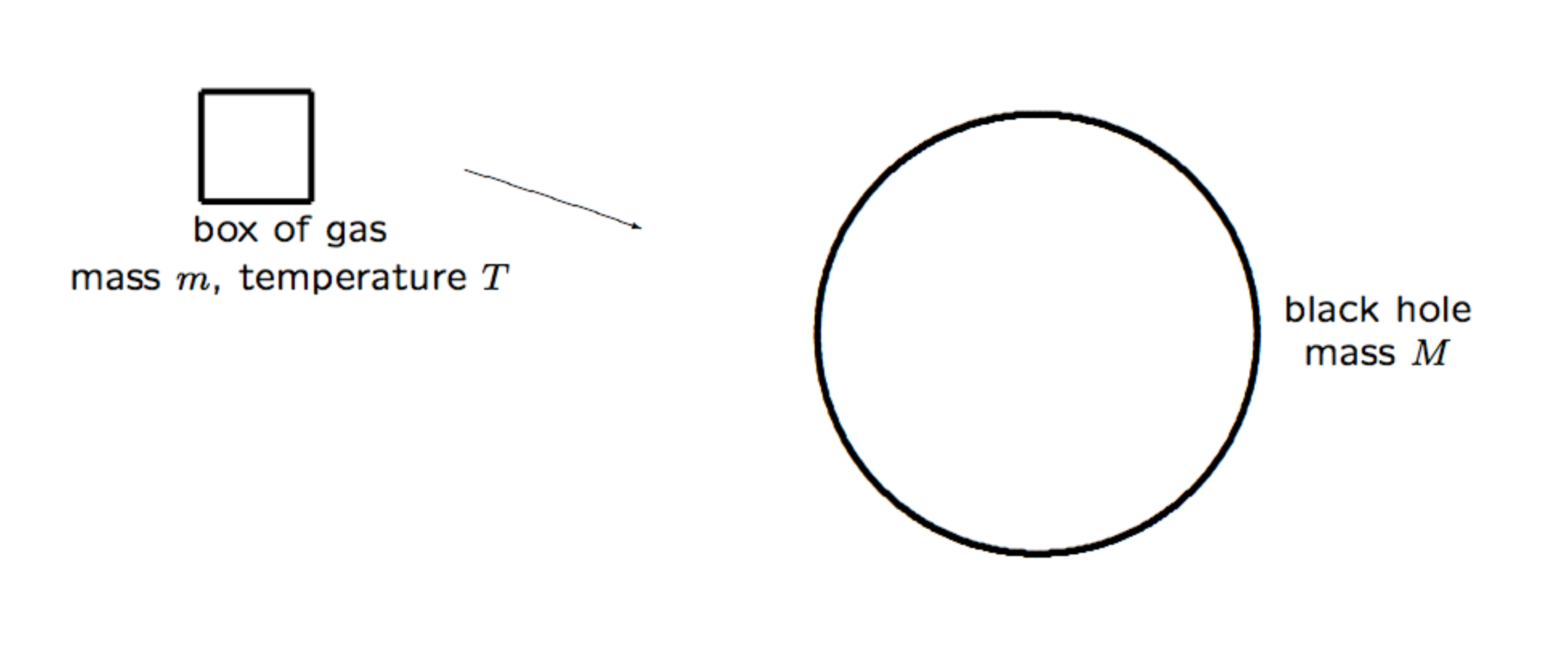}}
\caption{A thought experiment for Bekenstein-Hawking entropy}
\label{Carlipfig1}
\end{figure}

The box of gas will coalesce with the black hole when its proper
distance $\rho$ from the horizon is of order $\ell$.  For a Schwarzschild 
black hole, this proper distance is
\begin{align*}
\rho = \int_{2GM}^{2GM + \delta r} \frac{dr}{\sqrt{1- 2GM/r}}
 \sim \sqrt{GM\delta r}  ,
\end{align*}
so $\rho \sim \ell$ when $\delta r \sim \ell^2/GM$.  The gas initially 
has mass $m$, but its energy as seen from infinity is red-shifted 
as the box falls toward the black hole; when the box reaches 
$r = 2GM + \delta r$, the black hole will gain a mass 
\begin{align*}
\Delta M \sim m \sqrt{1 - \frac{2GM}{2GM + \delta r}} 
 \sim \frac{m\ell}{GM}  .
\end{align*}
The change in the horizon area is thus
\begin{align}
\Delta A \sim G^2M\Delta M \sim Gm\ell  .
\label{Carlipd2}
\end{align}
Comparing (\ref{Carlipd1}) and (\ref{Carlipd2}), we see that the black hole
must gain an entropy
\begin{align}
\Delta S_{\scriptscriptstyle\mathrm{BH}} \sim -\Delta S_{\mathrm{\scriptstyle gas}}
\sim \frac{\Delta A}{\hbar G}   .
\label{Carlipd3}
\end{align}

\subsection{Temperature \label{Tempa}}
We next estimate the Hawking temperature, at a similar level of hand-waving
\cite{Schutz}.  As in section 3, we consider inequivalent vacua,
but following a heuristic approach suggested by Hawking \cite{Hawkingc}, we 
now compare vacua inside and outside the horizon. 
\begin{figure}
\centerline{\includegraphics[width=4in]{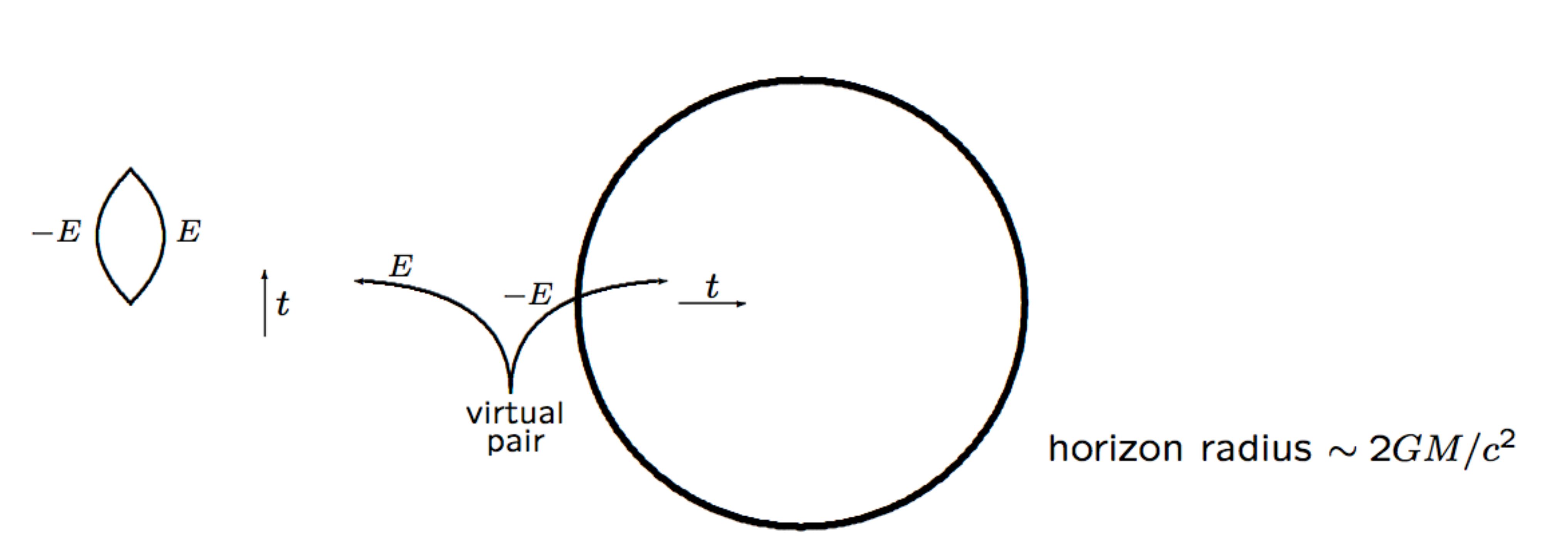}}
\caption{A thought experiment for Hawking temperature}
\label{Carlipfig2}
\end{figure}

As usual in quantum field theory, let us view the vacuum as a sea
of virtual pairs of particles, each pair consisting of a particle
with positive energy $E$ relative to an observer at infinity and a particle
with negative energy $-E$.  Normally, by
the uncertainty principle, a negative energy particle can exist only for a 
time $t\sim \hbar/|E|$.  Recall, however, that the distinction between 
``positive'' and ``negative'' energy can depend on the choice of time.
For a Schwarzschild black hole, with a metric
\begin{align}
ds^2 = \left(1-\frac{2M}{r}\right)dt^2 - \left(1-\frac{2M}{r}\right)^{-1}dr^2
        - r^2d\Omega^2  ,
\label{Carlipd4}
\end{align}
the coordinate $t$ is a time coordinate in the exterior, but a spatial coordinate
in the interior, where the coefficient of $dt^2$ changes sign.\footnote{Strictly 
speaking, the coordinates labeled $r$ and $t$ outside the 
horizon are not the same as those inside, since the two coordinate patches 
do not overlap.  The argument can be rephrased in terms of proper times of
infalling observers, though, in a way that avoids this issue.}
Similarly, the coordinate $r$ is a spatial coordinate in the exterior, but a time 
coordinate in the interior, where ``forward in time'' means ``towards the 
center of the black hole.''   A particle with negative energy relative to an 
exterior observer may thus have positive energy relative to an interior 
observer, and if it crosses the horizon quickly enough, this will allow us to 
evade the uncertainty principle.

Now consider a virtual pair of massless particles momentarily at rest at a 
coordinate distance $\delta r$ from the horizon.    
As in the preceding section, the proper distance --- and therefore the 
proper time for the negative energy partner to reach the 
horizon --- will be
\begin{align*}
\tau \sim \sqrt{GM\delta r} .
\end{align*}
Setting this equal to the lifetime $\hbar/E$ of the pair, we find 
that
\begin{align*}
|E| \sim \frac{\hbar}{\sqrt{GM\delta r}}  .
\end{align*}
which should match the energy of the escaping positive-energy 
partner.  This is the energy at $2GM + \delta r$, though.  As in
the preceding section, the energy at infinity will be red-shifted, becoming
\begin{align}
E_\infty \sim 
  \frac{\hbar}{\sqrt{GM\delta r}}\sqrt{1 - \frac{2GM}{2GM+\delta r}}
  \sim \frac{\hbar}{GM}  ,
\label{Carlipd5}
\end{align}
independent of the initial position $\delta r$.  We thus 
expect a black hole to radiate with a characteristic temperature 
$kT\sim\hbar/GM$, matching the Hawking temperature (\ref{Carlipa1}).

\section{The many derivations of black hole thermodynamics \label{Manya}}

Hawking's derivation of the black hole temperature seems to depend 
only on simple properties of quantum field theory in curved spacetime.  
Still, one may worry that these properties have been pushed beyond their 
realm of validity.  In particular, it is evident from section 2 that the
calculation of the Bogoliubov coefficients requires that we follow outgoing
modes backwards in time to a region in which they are very highly
blue-shifted, with energies far above the Planck scale 
\cite{Unruhb,Jacobsonb,Helfer}.

This ``trans-Planckian problem'' has been addressed in a number of ways: 
by imposing high-energy cut-offs \cite{Jacobsonb}, by restricting fields to 
discrete lattices \cite{Corley}, by altering dispersion relations to break the 
connection between high frequencies and high energies \cite{Jacobsonc,Schutzhold},
and by considering analog systems such as Unruh's ``dumb holes,'' fluid
flows that generate sonic event horizons \cite{Unruhb}.  It turns out to be
very hard to ``break'' Hawking's results: standard thermal  
Hawking radiation, with the usual Hawking temperature, appears even in
models that never involve Planck scale physics\cite{BLV}.

Perhaps the strongest evidence for the reality of black hole thermodynamics,
though, comes from the number of independent derivations, each relying on
different assumptions and approximations.  In this section I will briefly 
summarize some of these results.

\subsection{Other settings \label{othera}}

Hawking's original derivation involved a particular physical setting, a star
collapsing to form a black hole.  This determined the choice of 
vacuum state, and one could worry that the result might be too narrow.  
It has been subsequently shown, though, that one can generalize the
derivation considerably, for instance by replacing the initial Minkowski 
vacuum in the distant past with the vacuum for a freely falling observer 
near the horizon \cite{Unruhc}.  Indeed, there is now fairly strong evidence 
that only a few fundamental features are needed to obtain thermal 
radiation: a vacuum near the horizon as seen by a freely falling observer, 
vacuum fluctuations that start in the ground state, and subsequent 
adiabatic evolution \cite{Schutzhold,Visser}.  In particular, the Einstein
field equations are not required \cite{Visser_essential}.

We may also ask which ingredients in the derivation of section 
2 were essential to the result \cite{BLSV}.  Consider any
setting in which, as in figure \ref{Carlipfig0b}, one can trace null
geodesics (such as $\gamma$ in the figure) from part of $\mathcal{I}^+$
to part of $\mathcal{I}^-$, defining a map $p(u)$.  One can then 
\emph{define} a time-dependent ``surface gravity'' $\kappa(u)$ by
\begin{align}
\kappa(u) = -\frac{p^{\prime\prime}(u)}{p'(u)}  .
\label{Carlipe0}
\end{align}
Now pick a point $u_*$ on $\mathcal{I}^+$.  Barcel{\'o} {\it et al.}\
show \cite{BLSV} that an observer near $u_*$ will observe
nearly thermal Hawking radiation, with a time-dependent temperature
$\hbar\kappa(u_*)/2\pi$, provided an adiabatic condition
\begin{align} 
\kappa'(u_*)\ll \kappa(u_*)^2
\label{Carlipe0a}
\end{align}
holds.  In particular, it is not even necessary for an event horizon to
form; the exponential ``peeling'' (\ref{Carlipc9}) is sufficient.

One can also analyze an eternal black 
hole at equilibrium with its Hawking radiation \cite{HartleHawking},
obtaining consistent results.  Moreover, the computation may
be extended beyond the number operator (\ref{Carlipc6}) to find an 
expression for the full final state in terms of the initial vacuum; 
it is exactly thermal, at least in the approximation that one ignores 
back-reaction of Hawking radiation on the black hole \cite{Wald,Parkerb}.

\subsection{Unruh radiation \label{Unruha}}

By the principle of equivalence, the gravitational field near a black hole
horizon should be locally equivalent to uniform acceleration in a flat
spacetime.  Hawking radiation is not entirely local, so one might not
expect an exact equivalence, but one should be able to test Hawking's
results by comparing the vacua of an inertial observer and a uniformly
accelerating observer in Minkowski space.  This was first done by Unruh
\cite{Unruhc}.  The Bogoliubov coefficients can be derived fairly easily,
and the conclusion is that the accelerated observer with a constant proper 
acceleration $a$ sees a thermal flux of particles with a temperature
\begin{align}
T_U = \frac{\hbar a}{2\pi}  ,
\label{Carlipe1}
\end{align}
in almost exact analogy with the Hawking temperature (\ref{Carlipa1}).

Coincidentally, at almost the same time as Unruh's work, Bisognano 
and Wichmann gave an independent, more difficult but mathematically 
rigorous proof of the Unruh effect in quantum field theory \cite{Bisognano}.
Their results are quite general; while they require flat spacetime,
the quantum field theory need not be a free theory.

\subsection{Particle detectors \label{detecta}}

Hawking's derivation and similar approaches to black hole thermodynamics 
depend heavily on the standard quantum field theoretical definitions of
vacuum states, Fock space, and particle number.  But since one of the 
main conclusions is that states in curved spacetime are rather different 
from those we are used to in Minkowski space --- for instance, the vacuum 
is no longer unique --- we might worry that we are overinterpreting 
the formalism.  True physical observables are notoriously difficult to find
in quantum gravity, so this is not a trivial concern.

To investigate this question, Unruh \cite{Unruhc} and DeWitt \cite{DeWitt}
developed simple models of particle detectors in black hole backgrounds.
They showed that such detectors do, in fact, see thermal radiation at the
Hawking temperature.  Similarly, Yu and Zhou \cite{Yu}  have shown that 
a two-level atom outside a black hole will spontaneously excite as if it
were in a thermal bath at the Hawking temperature.

\subsection{Tunneling \label{Tunna}}

Hawking radiation is a quantum process, and we may try to apply our
intuition about quantum mechanics to seek a deeper understanding. 
As noted in section 3, an obvious guess is that while escape 
from a black hole is classically forbidden, quantum particles might
tunnel out.  For instance, instead of visualizing virtual pairs forming
outside the horizon as in section 2, we might reverse the 
picture and imagine pairs forming inside the horizon, with 
one member then tunneling out.

While the idea of a tunneling description can be traced to Damour and
Ruffini \cite{Damour} and Srinivasan and Padmanabhan \cite{SriPad}, 
its modern incarnation owes itself to a key insight
of Parikh and Wilczek \cite{Parikh}, who realized that rather than 
thinking of a particle as tunneling through the horizon, one could think 
of the horizon as tunneling past the particle.  
Consider an initial Schwarzschild black hole of mass $M$ that emits
a particle in an s-wave state, represented as a spherical shell of mass
$\omega \ll M$.  In the WKB approximation, the tunneling rate is
\begin{align}
\Gamma = e^{-2\mathop{Im} I/\hbar}  ,
\label{Carlipe3}
\end{align}
where $I$ is the action for the outgoing shell, moving in a background
metric of a black hole as it crosses the horizon from
$r_{\scriptscriptstyle\mathit{in}}$ to $r_{\scriptscriptstyle\mathit{out}}$.
The exponent, in turn, can be written as
\begin{align}
\mathop{Im} I = 
\mathop{Im}
\int_{r_{\scriptscriptstyle\mathit{in}}}^{r_{\scriptscriptstyle\mathit{out}}}
p_r dr = 
\mathop{Im}
\int_{r_{\scriptscriptstyle\mathit{in}}}^{r_{\scriptscriptstyle\mathit{out}}}
\int_0^{p_r}dp_r'\,dr =
\mathop{Im}\int_M^{M-\omega} 
\int_{r_{\scriptscriptstyle\mathit{in}}}^{r_{\scriptscriptstyle\mathit{out}}}
\frac{dr}{{\dot r}}dH ,
\label{Carlipe4}
\end{align}
where I have used Hamilton's equations of motion to write 
$dp_r = dH/{\dot r}$ and noted that the horizon moves inward from 
$GM$ to $G(M-\omega)$ as the particle is emitted.

To evaluate this integral, Parikh and Wilczek use the Painlev{\'e}-Gullstrand
form of the black hole metric,
\begin{align}
ds^2 = \left(1-\frac{2G(M-\omega)}{r}\right)dt^2 
  - 2\sqrt{\frac{2G(M-\omega)}{r}}dt\,dr - dr^2 - r^2d\Omega^2  ,
\label{Carlipe2}
\end{align}
which avoids coordinate singularities at the horizon.  Outgoing radial 
null geodesics then satisfy
\begin{align*}
{\dot r} = 1 - \sqrt{\frac{2G(M-\omega)}{r}} 
\end{align*}
and the integral  (\ref{Carlipe4}) can be evaluated by a standard contour
deformation, yielding a tunneling rate
\begin{align}
\Gamma = e^{-8\pi\omega G\left(M - \frac{\omega}{2}\right)/\hbar} 
 = e^{\Delta S_{\scriptscriptstyle\mathrm{BH}}}
\label{Carlipe5}
\end{align}
where $\Delta S_{\scriptscriptstyle\mathrm{BH}}$ is the change in the 
Bekenstein-Hawking entropy (\ref{Carlipa1}) due to the emission of the shell.  
By the first law of thermodynamics, $\Delta S_{\scriptscriptstyle\mathrm{BH}} 
= -\hbar\omega/T_{\scriptscriptstyle\mathrm{H}}$, and we recover the 
expected emission rate for thermal Hawking radiation.  

The tunneling approach was initially developed for the simplest black holes,  
but has by now been vastly extended, to apply to virtually every known
black hole solution, including even nonstationary configurations 
\cite{Vanzoreview}.  The method has also been generalized; in particular,
one can replace the assumption that outgoing particles follow null 
geodesics by the more general approximation that the action $I$ satisfies 
a Hamilton-Jacobi equation \cite{HJ}.   

\subsection{Hawking radiation from anomalies \label{anoma}}

From early on, it was clear that black hole thermodynamics should be
visible in the behavior of the stress-energy tensor near a   
horizon \cite{DeWitt,Daviesx}.  The computation of the expectation
value $\langle T_{ab}\rangle$ is a complicated and technical subject, 
beyond the scope of this review; see, for instance, the book by Birrell and 
Davies \cite{BirrellDavies} for a pedagogical introduction.  The results, however, 
are consistent with Hawking's discovery.  In particular, the heuristic approach
of section 2 is supported:  an ingoing flux of negative energy 
near the horizon balances the outgoing flux of positive energy at infinity.  
Energy is conserved, and as in the tunneling derivation of section 4, 
the back-reaction of Hawking radiation reduces the mass of the black hole.

In general, the computation of  the expectation value $\langle T_{ab}\rangle$
requires a mixture of analytic approximations and numerical methods.  
In one special case, though, the situation becomes drastically simpler.  
Consider a conformally invariant field (for instance, a massless scalar)
in two spacetime dimensions.  Classically, conformal invariance forces
the trace $T^a{}_a$ of the stress-energy tensor to vanish, and its quantum 
correction, the ``trace anomaly,'' is completely determined as
\begin{align*}
\langle T^a{}_a\rangle = \frac{c}{24\pi} R  ,
\end{align*}
where $c$, the central charge, is fully determined by the properties of the 
field in a flat background \cite{FMS}.  Christensen and Fulling \cite{Christensen}  
realized in 1977 that in such a case, the full stress-energy tensor  is 
completely determined  from $\langle T^a{}_a\rangle$ by conservation 
laws and boundary conditions.   The result exactly matched expectations,
with fluxes of positive energy at infinity, describing thermal radiation, 
balanced by fluxes of negative energy through the horizon.

Christensen and Fulling's original analysis relied crucially on the
restriction to two dimensions.  More recently, though Wilczek and   
collaborators \cite{Robinson,Iso,Umetsu} have shown how to generalize
the argument to a much wider setting.  The central idea is that one can
impose an effective description of a vacuum state near the horizon,
the equivalent of the freely falling observer's vacuum, by excluding
matter fields that correspond to outgoing (``horizon-skimming'') modes.  
This restriction makes the field theory chiral, though, and a chiral theory
contains a new anomaly, a diffeomorphism anomaly.  By analyzing the 
resulting theory one partial wave at a time, one can again use this anomaly 
to reconstruct the stress-energy tensor, recovering the characteristic 
thermal radiation.  This method is actually closely related to that
of Christensen and Fulling
--- in two dimensions, one can always trade a conformal anomaly for a 
diffeomorphism anomaly by adding appropriate counter\-terms --- but
the newer approach seems more suitable for treating separate partial
waves, and thus for application to more than two dimensions.

In a remarkable piece of work, Iso, Morita, and Umetsu \cite{IMU}
and Bonora, Cvitan, Pallua, and Smoli{\'c} \cite{BCa,BCb}  have
extended this approach to obtain a much more detailed
picture of Hawking radiation. They show that if one considers not
 just the stress-energy tensor, but higher spin currents as well, 
one can recover the full thermal spectrum.
Hawking radiation can thus be obtained, with only
fairly minimal assumptions, from the symmetries of the near-horizon 
region of a black hole.

\subsection{Periodic Greens functions \label{Greena}}

In ordinary quantum field theory,
thermal Greens functions have a hidden periodicity.  Consider, for
instance, the Greens function of a scalar field $\varphi$ in a thermal
ensemble at temperature $T = 1/\beta$:
\begin{align}
G_\beta(x,0;x',t) 
    &= \mathop{Tr}\left(e^{-\beta H}\varphi(x,0)\varphi(x',t)\right) 
    = \mathop{Tr}\left(e^{-\beta H}\varphi(x,0)e^{-\beta H}e^{\beta H}%
      \varphi(x',t)\right)\nonumber\\
   &=\mathop{Tr}\left(\varphi(x,0)e^{-\beta H}e^{\beta H}\varphi(x',t)%
  e^{-\beta H}\right) 
 =\mathop{Tr}\left(\varphi(x,0)e^{-\beta H}\varphi(x',t+i\beta)\right)\nonumber\\
 &= G_\beta(x',t+i\beta;x,0) \label{Carlipe6}
\end{align}
where I have used cyclicity of the trace and the fact that the 
Hamiltonian generates time translations, so 
$e^{\beta H}\varphi(x',t)e^{-\beta H} = \varphi(x',t+i\beta)$.  
In particular, a thermal Greens function that is symmetric in its arguments 
must be periodic in imaginary time, with period $i\beta$.  Conversely,
in axiomatic quantum field theory such periodicity, formalized
as the KMS condition \cite{Kubo,Martin,Haag}, serves to define
a thermal state.

For a uniformly accelerating observer in Minkowski space, it was
shown in 1976 that the Greens function has just such a periodicity
\cite{Bisognano}, with a temperature equal to the Unruh temperature
(\ref{Carlipe1}).  By the arguments of section 2, one
should expect an analogous result for a stationary observer
near a black hole horizon.  This is indeed the case \cite{Gibb,Gibc},
and  $\beta$ is just the inverse of the
Hawking temperature (\ref{Carlipa1}).  Moreover, the response of
a particle detector of the sort discussed in section 3
is determined by a closely related Greens function, and the
periodicity ensures quite generally that such a detector will see
a thermal bath at temperature $T_{\scriptscriptstyle\mathrm{H}}$.

\subsection{Gravitational partition function \label{parta}}

One may ask whether the periodicity properties of the Greens
functions may be extended to the gravitational field itself.
For ordinary quantum mechanical systems, it is well known that 
a thermal partition function may be obtained from the
usual path integral by analytically continuing to periodic
imaginary time with period $\beta$ \cite{Feynman}.  The meaning
of such a continuation is not so clear in general
relativity, though: there is usually no preferred time coordinate to 
make imaginary, and there is no simple relationship between
solutions of the field equations with ``real time'' (Lorentzian
signature) and ``imaginary time'' (Riemannian signature).  

In one setting, though, there is a natural choice.  If a spacetime 
is stationary --- that is, if it admits a timelike Killing vector --- 
the Killing vector defines a preferred time coordinate.  In particular, 
for a nonextremal stationary black hole, the generic metric near 
enough to the horizon takes the form
\begin{align*}
ds^2 = 2\kappa(r-r_+)dt^2 - \frac{1}{2\kappa(r-r_+)}dr^2 
  - r_+^2d\Omega^2   ,
\end{align*}
where $r_+$ is the location of the horizon and $\kappa$ is again 
the surface gravity.  Setting $t=i\tau$ and replacing $r$ by the 
proper distance to the horizon,
\begin{align*}
\rho = \frac{1}{\kappa}\sqrt{2\kappa(r-r_+)}  ,
\end{align*}
we obtain the ``Euclidean black hole'' metric
\begin{align}
ds^2 = d\rho^2 + \kappa^2\rho^2d\tau^2  + r_+^2d\Omega^2 .
\label{Carlipe7}
\end{align}

We can immediately recognize the first two terms of (\ref{Carlipe7})
as the metric of a flat plane in polar coordinates.  The 
horizon $r=r_+$ has collapsed to a point $\rho=0$: the 
``Euclidean section'' includes only the black hole exterior.  
But the $\rho$--$\tau$ plane is flat only if $\kappa\tau$ has a 
period $2\pi$; otherwise, the origin is a conical singularity, and the 
metric will not extremize the action at that point.  This periodicity, 
in turn, requires that $\tau$ have a period 
$2\pi/\kappa = 1/T_{\scriptscriptstyle\mathrm{H}}$,
exactly as in the preceding section.

We now see that the periodicity of Greens functions did not depend 
on details of quantum fields, but arose directly from the geometry.  But 
we can do more: as Gibbons and Hawking argued \cite{GibHawk}, 
we can use this result to obtain a saddle point approximation to the 
gravitational path integral.

By analogy with ordinary quantum field theory,
the Euclidean path integral for the gravitational partition function can 
be formally written as
\begin{align}
Z[\beta] = \int [dg] e^{I_{\scriptscriptstyle\mathrm{Euc}}}
\label{Carlipe8}
\end{align}
where $I_{\scriptscriptstyle\mathrm{Euc}}$ is the ``Euclidean'' 
Einstein-Hilbert action --- that is, the Einstein-Hilbert action for
metrics of Riemannian signature --- and the integral is over all
metrics that are, in some sense, periodic in ``Euclidean'' 
time with period $\beta$.  This is, of course, an ill-defined expression: 
general relativity is nonrenormalizable, so even if we knew exactly what 
``periodic'' meant here, we would not know how to make sense of the 
functional integral.  
Still, though, there ought to be some sense in which (\ref{Carlipe8})  
has a meaning in effective field theory \cite{Burgess}, and even if the
full expression is ill-defined, a saddle point approximation could
give a physically reasonable result. 

Naively, the saddle point contributions (\ref{Carlipe7}) 
gives a vanishing contribution to
$I_{\scriptscriptstyle\mathrm{Euc}}$: the ``bulk'' Einstein-Hilbert action
\begin{align*}
\frac{1}{16\pi G}\int d^4x\,\sqrt{|g|}R
\end{align*}
is zero for any classical solution of the vacuum field equations.  The  
key observation of Gibbons and Hawking \cite{GibHawk} was that on
a manifold with boundary, the ``bulk'' action must be supplemented 
by a boundary term, without which the action will  typically have no true
extrema \cite{Regge}.   The boundary was originally taken to be
at infinity, but subsequent work has shown that it may also be
placed at the horizon \cite{BTZa,Teitelboim,HawkingHorowitz}.  
The resulting boundary term may be evaluated in a number of ways;
for instance, if one dimensionally reduces the action to the 
$\rho$--$\tau$ plane, the problem becomes purely topological 
\cite{BTZa}.  The final result is that
\begin{equation}
I_{\scriptscriptstyle\mathrm Euc} 
 = \frac{\ A_{\mathrm{\scriptstyle hor}}}{4\hbar G}
 - \beta(M + \Omega J + \Phi Q) .
\label{Carlipe9}
\end{equation}
The saddle point contribution $e^{I_{\scriptscriptstyle\mathrm{Euc}}}$ 
to the partition function (\ref{Carlipe8}) may then be recognized as the 
grand canonical partition function for a system with inverse temperature 
$\beta = 2\pi/\kappa$ and entropy $S_{\scriptscriptstyle\mathrm{BH}}%
=  A_{\mathrm{\scriptstyle hor}}/4\hbar G$, just as expected.  

The same analysis applies to much more general stationary geometries 
\cite{HawkingHunter}.  Just as above, Killing horizons in the Lorentzian 
configurations translate into zeros of the Killing vector in the Riemannian 
continuation, and boundary terms resembling (\ref{Carlipe9}) again appear.   
Recently Neiman has argued that even in Lorentzian signature,
the usual Gibbons-Hawking boundary term acquires an imaginary 
piece whenever the boundary involves ``signature flips,'' 
points at which the signature of the \emph{boundary} changes 
from spacelike to timelike \cite{Neiman}.  For 
black hole spacetimes, the resulting boundary term looks very 
much like the one in the Euclidean action.
 
Alternatively, one may try to evade the ambiguities in ``imaginary 
time'' by starting with the Hamiltonian formulation of general relativity 
\cite{Teitelboim}.   Perhaps unsurprisingly, the same boundary term
(\ref{Carlipe9}) appears.  In canonical quantization, this boundary term  
gives rise to a new term in the Wheeler-DeWitt equation, from which 
one can again recover the Bekenstein-Hawking entropy \cite{CarTeit}.

\subsection{Pair production of black holes}

As Heisenberg and Euler noticed in 1936 \cite{Heisenberg}, the quantum
vacuum is unstable in the presence of a strong electric field.  Virtual 
pairs of charged particles can be ``pulled out of the vacuum'' by the
field, becoming physical particle-antiparticle pairs.  The pair production
rate per unit volume for a fermion of mass $m$ and charge $e$ in a 
constant field $\mathcal{E}$ was worked out in detail by Schwinger 
\cite{Schwinger}, and takes the form
\begin{align*}
W \sim \frac{\alpha^2\mathcal{E}^2}{\pi^2} e^{- \pi m^2/|e\mathcal{E}|}  .
\end{align*}
For $\mathcal{N}$ identical species of fermions, this expression would be
multiplied by $\mathcal{N}$.

Now consider pair production of charged black holes in an electric field.  
Without a full quantum theory of gravity we cannot compute an exact 
production rate, but as in the preceding section, we can use the
Euclidean path integral  to obtain a semiclassical approximation.  
If the Bekenstein entropy is a statistical mechanical entropy, we would 
expect a black hole to have $\exp\{S_{\scriptscriptstyle\mathrm{BH}}\}$
microscopic states, with a corresponding enhancement of the pair
production rate by $\mathcal{N} = \exp \{S_{\scriptscriptstyle\mathrm{BH}}\}$.

This is exactly what is found.  The first computation, by Garfinkle,
Giddings, and Strominger \cite{Garfinkle}, considered magnetically
charged black holes, and compared the pair production rate to the
rate for magnetic monopoles of the same mass.  Subsequent
work extended these results to a much wider variety of black holes
\cite{Brown,Dowker,MannRoss}.  In every known case, pair production
is enhanced just as one would expect if the Bekenstein-Hawking entropy
represented an actual counting of states.

\subsection{Quantum field theory and the eternal black hole \label{eterna}}

Hawking's derivation of black hole thermodynamics was based on a 
collapsing star, with a Minkowski-like vacuum
in the distant past.  But we can also write down solutions of the Einstein
field equations that represent eternal black holes.   
\begin{figure}
\centerline{\includegraphics[width=2.6in]{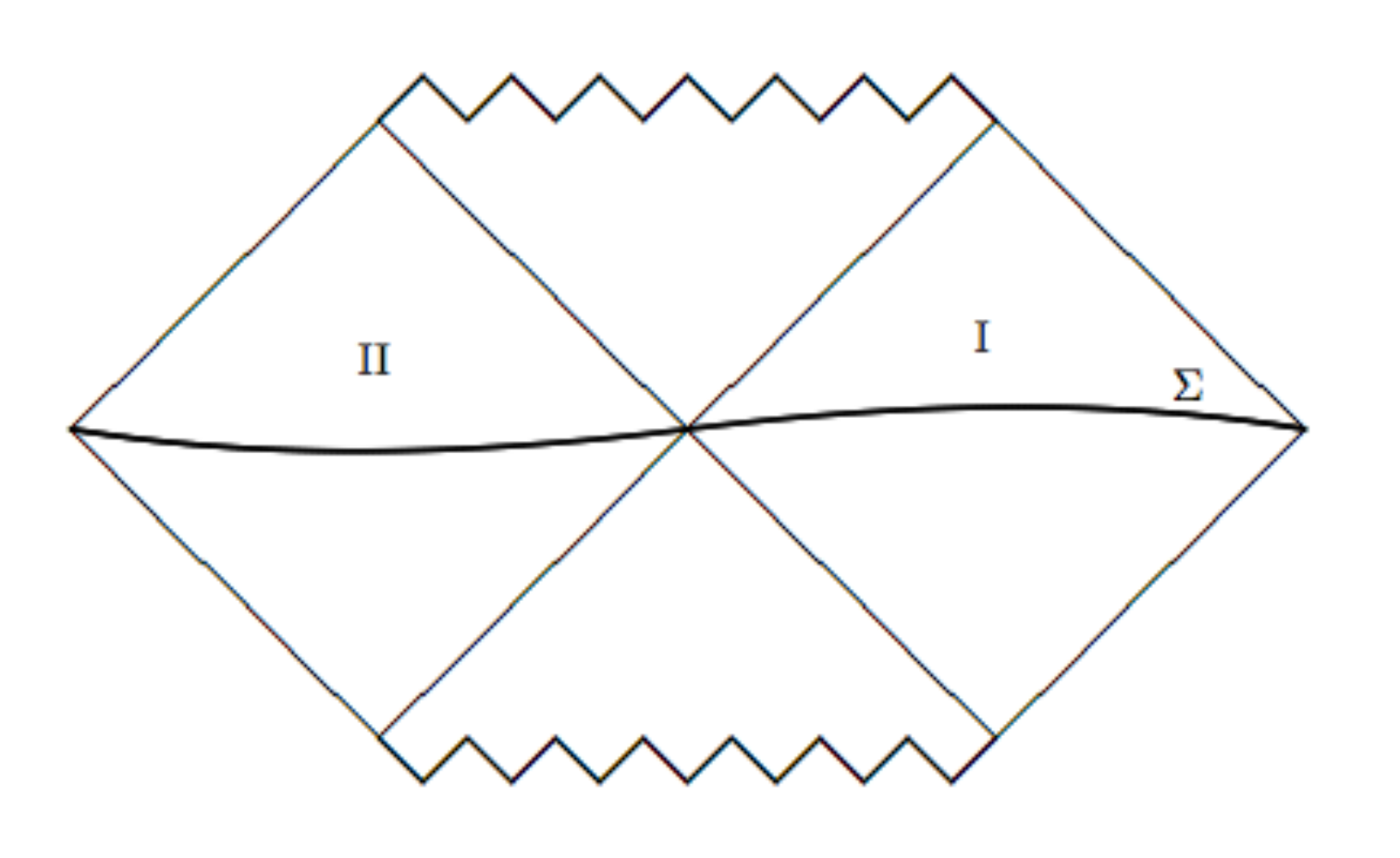}}
\caption{A Carter-Penrose diagram for an eternal black hole}
\label{Carlipfig3}
\end{figure}

Recall that the
maximally extended eternal black hole spacetime has a Carter-Penrose 
diagram with four regions, as shown in figure \ref{Carlipfig3}.
Now consider a quantum state $\Psi$ defined on a Cauchy surface 
$\Sigma$ passing through the black hole bifurcation sphere, as shown 
in the figure.  No information can pass from region II to region I, so even 
if $\Psi$ is a pure state, an observer in region I will see only a density 
matrix, obtained by tracing over region II.  This at least opens up the
possibility that physics for an observer in region I will be thermal.

Indeed, for a free quantum field there is at most one quantum state, 
the Hartle-Hawking vacuum state, that is regular everywhere on the 
horizon \cite{Waldbk,Kay}.   For a scalar field, a direct computation 
shows that the density matrix obtained by tracing over region II 
is thermal, with a temperature $T_{\scriptscriptstyle\mathrm{H}}$ 
\cite{Israelb}.  For more general fields, the same can be shown by 
means of fairly sophisticated quantum field theoretical arguments
\cite{Waldbk,Kay} or by an analysis of the path integral \cite{Jacobsond}.
Thus even if matter is secretly in a pure state, the eternal black hole is 
effectively also a thermal system.

\subsection{Quantized gravity and classical matter \label{BTZa}}

Although they differ in detail, the derivations of Hawking radiation 
so far have all analyzed quantum matter in a classical 
black hole background.  This is a reasonable approximation,
except perhaps for the ``trans-Planckian problem'' discussed
above.  In one special case, though, the direction can be reversed:
Hawking radiation can be derived from classical matter in the
presence of a quantum black hole.

This special case is that of the BTZ black hole \cite{BTZ}, a black
hole in (2+1)-dimensional asymptotically anti-de Sitter space.
This is a rather peculiar solution: the vacuum Einstein field equations
in 2+1 dimensions imply that spacetime has constant negative
curvature, and, indeed, the BTZ black hole can be described as a
region of anti-de Sitter space with certain identifications \cite{BHTZ}.
Nevertheless, it is a genuine black hole, the final state of collapsing
matter, with a normal event horizon and (in the rotating case) a
normal inner Cauchy horizon \cite{CarlipBTZ}.

The BTZ black hole presents a puzzle for any statistical mechanical
explanation of black hole entropy.  As a corollary to the constant curvature
required by the field equations, (2+1)-dimensional gravity contains 
no propagating degrees of freedom that could serve as candidates for
microscopic states.  As first suggested by Carlip \cite{Carlipedge} and
confirmed in detail by Strominger\cite{Strominger} and, independently,
Birmingham, Sachs, and Sen \cite{BSS}, the natural candidate for the
microscopic degrees of freedom are ``edge states,'' additional degrees
of freedom at infinity that appear because the boundary conditions
break diffeomorphism invariance.   

I will discuss these states  further in section 9.  For now, 
the key feature is that they allow an ``effective'' description of the 
quantum BTZ black hole in terms of a particular two-dimensional conformal
field theory at infinity \cite{CarlipBTZb}.   Emparan and Sachs have shown 
how to couple this quantum theory to classical matter \cite{ES}, and 
to use the coupling to calculate transition amplitudes between quantum 
black hole states induced by interactions with this ``background'' matter.  
Detailed balance arguments then allow them to determine the emission 
rate of the matter fields, obtaining the correct Hawking radiation spectrum, 
even including the correct greybody factors.

\subsection{Other approaches \label{otherb}}

I have tried to give an overview of the most common approaches to
black hole thermodynamics, but I have necessarily omitted a few.  
For example:

\begin{itemize}
\item Instead of considering only  the region outside the horizon, 
one may choose a time-slicing that crosses the horizon and continues 
to the singularity.  A black hole interior is not static --- the Killing 
vector that gives time translation invariance in the exterior 
becomes spacelike in the interior --- and the Hamiltonian in 
horizon-crossing coordinates is therefore time-dependent.
Nevertheless, one can perform a Heisenberg picture quantization to
obtain time-dependent states which, although pure, behave
``thermally'' in the sense that they excite radiation detectors and have
a net flux of radiation at infinity that matches Hawking's results
\cite{Melnikov}.  

\item One may repeat the computation of Unruh radiation in the functional
Schr{\"o}dinger picture, in which the quantum state is a wave functional
of the field configuration.   The results are again what would be expected 
\cite{Freese}: the ground states for the stationary and accelerated 
observers are different, and the difference appears as a thermal
distribution of ``Minkowski'' particles in the accelerated observer's
state. 

\item York has proposed a model in which fluctuations of 
the event horizon lead to a ``quantum ergosphere'' through which
particles can tunnel \cite{York}.  The picture is too complicated
to  compute an exact Hawking temperature, but by 
restricting the horizon fluctuations to those corresponding to the
lowest quasinormal mode, he obtains a value for the temperature 
that is accurate to to within about 2\%.
\end{itemize}

The derivations  
in this section are not  entirely independent.  For instance, the 
response function of a particle detector (section 3)
depends on a Greens function, and it is the periodicity of 
this function (section 6) that determines the thermal response.   
This periodicity, in turn, may be traced the periodicity 
of the black hole metric in imaginary time (section 7).  The 
tunneling approach (section 4) and the original analysis
by Hawking both lead to relations of the form
\begin{align*}
P(\mathrm{emis}) = e^{-\beta E}P(\mathrm{absor})
\end{align*}
between emission and absorption probabilities, and
both results come from the behavior of complex functions
near the horizon.   Still, the derivations
are sufficiently independent, with different enough assumptions and
approximations, that together they provide very powerful evidence for
the reality of Hawking radiation.

\section{Thermodynamic properties of black holes}

There is more to thermodynamics than temperature and entropy.
If black holes are thermal  objects, we should be able to apply
the rest of thermodynamic theory, from Carnot cycles
to phase changes, to systems containing black holes.  
In this section, after a short discussion of black hole evaporation,
I will review four selected aspects: heat capacity, 
phase transitions, thermodynamic volume, and an   
argument that black hole thermodynamics prohibits certain violations 
of Lorentz invariance.

\subsection{Black hole evaporation}

We have seen that a black hole radiates as a grey body, at a temperature 
$T_{\scriptscriptstyle\mathrm{H}}$.  By the Stefan-Boltzmann law, it
will therefore lose mass at a rate
\begin{align}
\frac{dM}{dt} = -\varepsilon\sigma T_{\scriptscriptstyle\mathrm{H}}^4
     A_{\scriptscriptstyle\mathrm{hor}}
\label{Carlipf0}
\end{align}
where $\varepsilon$ is some averaged emissivity, summed over the species
of particles that can be radiated.  For the Schwarzschild black hole, 
$T_{\scriptscriptstyle\mathrm{H}}\sim 1/M$ and 
$A_{\scriptscriptstyle\mathrm{hor}}\sim M^2$, so the lifetime 
goes as $M^3$.  To determine the exact coefficients one
must treat the greybody factors carefully \cite{Page}; the result is
that
\begin{align*}
\tau  \sim 10^{71}(M/M_\odot)^3
\end{align*}
where $M_\odot$ is the mass of the Sun.  For charged or rotating black
holes, evaporation is more complicated \cite{Hiscock}: black 
holes can ``discharge'' via Schwinger pair production, but the Hawking 
temperature decreases with increasing charge and angular 
momentum.

\subsection{Heat capacity}

A Schwarzschild black hole has a temperature $T = \hbar/8\pi M$.
As an isolated black hole evaporates, it radiates energy, its mass decreases,
and its temperature consequently \emph{increases}.  That is, a Schwarzschild 
black hole has a negative heat capacity,
\begin{align}
C = T\frac{\partial S}{\partial T} = -\frac{1}{8\pi T^2}  .
\label{Carlipf2}
\end{align}
For a rotating black hole, the situation is more complicated: as
Davies has shown \cite{Davies}, the heat capacity at fixed angular 
velocity is always negative, but the heat capacity at fixed angular
momentum is negative for low $J$, diverges as a critical value,
and then becomes positive, a behavior
indicative of a phase transition.  

While negative heat capacity is unusual,
it is not a peculiar feature of black holes, but holds even
for Newtonian self-gravitating systems.  As a satellite in
low Earth orbit loses energy to atmospheric friction, its orbit drops
to a lower altitude, \emph{increasing} its kinetic energy.  For 
gravitationally bound stellar systems, Lynden-Bell has analyzed 
this behavior, which he calls it the ``gravothermal catastrophe,''
in detail \cite{LyndenBell}.  In Newtonian gravity, the runaway 
behavior is a consequence of the purely attractive, long range 
nature of the gravitational interaction, and it stops when gravity
is no longer the principle interaction governing the system.  For
a relativistic black hole, on the other hand, there seems to be no
competing interaction, at least until the evaporating black hole 
has shrunk to the Planck scale.

There are two standard (theoretical) ways to stabilize a black hole.
The first is to place it in a reflecting cavity, allowing it to come
into equilibrium with its Hawking radiation.  For a spherical cavity 
of area $A = 4\pi r_B^2$, the heat capacity for a Schwarzschild 
black hole becomes \cite{Yorkb}
\begin{align}
C_A = 8\pi M^2 \left(1-\frac{2M}{r_B}\right)%
          \left(\frac{3M}{r_B} -1\right)^{-1}  ,
\label{Carlipf3}
\end{align}
which is positive for a small enough cavity, $r_B<3M$.   The second
is to consider a black hole in asymptotically anti-de Sitter space
\cite{HawkingPage}, in which the geometry serves as a sort of 
``cavity.''  For a given temperature, one finds two possible anti-de 
Sitter black hole configurations, a ``small'' black hole with negative 
heat capacity and a ``large'' black hole with positive heat capacity.

\subsection{Phase transitions}

As already indicated, black holes can have a complicated phase
structure, including phase transitions of various types.  The most famous
of these is the Hawking-Page transition for a black hole in asymptotically 
anti-de Sitter space \cite{HawkingPage}, a first order phase transition 
between thermal radiation at low temperatures and a black hole at 
higher temperatures.  In the AdS/CFT correspondence of string theory, 
this transition has an interesting ``dual'' description as a confinement-%
deconfinement transition \cite{Witten_Page}.

We have come to understand that this kind of behavior is not at all
exceptional.  As one example among many \cite{CarlipVaidya}, 
consider a black hole with charge $Q$ in asymptotically flat 
space, enclosed in a cavity whose walls are maintained at a fixed
temperature $1/\beta$.  One finds a phase diagram in the $Q$--$\beta$ 
plane containing a line of first-order phase transitions, terminating 
at a critical point which is the location of a second-order phase transition
with calculable critical exponents.

\subsection{Thermodynamic volume}

The first law of black hole mechanics (\ref{Carlipb1}) closely
resembles the standard form of the first law of thermodynamics,
but it is missing a term, the  pressure term $-pdV$.  There
is, in fact, a candidate for ``pressure'': the
cosmological constant $\Lambda$ has the correct dimension, and
acts as a pressure in cosmological settings.  But $\Lambda$ is not
 normally considered a state function; it is a coupling constant that
 is fixed by the theory, not by the state of the thermodynamic system.

There are, however, situations in which $\Lambda$ is a dynamical
field, forced to be constant only by virtue of its field equations
\cite{HennTeit1}.  In that case, it may make sense to treat it as
a state function \cite{Caldarelli}, and to generalize the first law 
to the form \cite{Sekiwa,KRT}
\begin{align}
\delta M = T\delta S + \Omega \delta J + \Phi \delta Q 
    + V_{\hbox{\scriptsize\it therm}}\delta p  ,
\label{Carlipf4}
\end{align}
where $p = - {\Lambda}/{8\pi G}$.  Note that in this formulation, 
the mass $M$ should be interpreted as enthalpy (that is, $U+pV$,
where $U$ is the internal energy) \cite{KRT}.

The ``thermodynamic volume'' $V_{\hbox{\scriptsize\it therm}}$ 
is defined by (\ref{Carlipf4}), and exists even in the limit 
$\Lambda\rightarrow0$.  Its physical meaning is not very well 
understood, though.  For a Schwarzschild black hole, it is easy 
to check that
\begin{align}
V_{\hbox{\scriptsize\it therm}} = \frac{4}{3}\pi r_+{}^3   ,
\label{Carlipf5}
\end{align}
where $r_+$ is the location of the horizon in Schwarzschild coordinates.
This relation generalizes to higher dimensional static black holes,
including those with charge: the thermodynamic volume is the volume
of a \emph{flat} round ball whose surface area is equal to the area of the 
event horizon.  But the generalization to rotating black holes is complicated, 
and the corresponding thermodynamic volume does not yet have a simple
physical interpretation \cite{Cvetic}.

The inclusion of a volume term in the first law (\ref{Carlipf4}) enriches
the phase structure of the theory, with a number of interesting
consequences \cite{Dolan,MannKub}.  For example, 
the equation of state for a charged, rotating anti-de Sitter black hole 
becomes very close to that of a van der Waals gas; rotating anti-de Sitter 
black holes in more than six dimensions exhibit reentrant phase 
transitions; and anti-de Sitter black holes in six dimensions with multiple 
spins have a triple point.

\subsection{Lorentz violation and perpetual motion machines}

Lorentz invariance guarantees that there is a single ``maximum speed''
$c$, the same for all particles.  I have implicitly used this assumption 
throughout this review, choosing units $c=1$.  Suppose, though,
that Lorentz invariance is broken, in such a way that two species of
particles --- say $A$ and $B$ --- have different maximum speeds
$c_B>c_A$.  Then Dubovsky and Sibiryakov have shown  \cite{DubSib} 
that a black hole can be used to build a ``perpetual motion machine
of the second kind,'' a device that transfers heat from a cold reservoir to 
a hot reservoir without any net use of energy.  In some 
Lorentz-violating theories, there is even a classical mechanism, 
analogous to the Penrose process, for ``mining'' a black hole \cite{Eling}. 

To understand the argument, first note that the event horizon for
the more slowly moving species $A$ will lie outside the event horizon 
for the faster species $B$.  This is physically intuitive, but also follows
directly if we restore the factors of $c$ in the definition of the
Schwarzschild event horizon, $r_+ = 2GM/c^2$.
Correspondingly, the Hawking temperature $T_{A,H}$ of species $A$ 
should be lower than the Hawking temperature $T_{B,H}$ of species $B$.  

Dubovsky and Sibiryakov's ``construction manual'' is then the 
following.  As heat reservoirs, construct a shell $A$ at temperature
$T_{A,\hbox{\scriptsize\it shell}}$ that interacts only with species $A$,
and a second shell $B$ at temperature $T_{B,\hbox{\scriptsize\it shell}}$ 
that interacts only with species $B$.  Arrange the temperatures so that
\begin{align}
T_{B,H}>T_{B,\hbox{\scriptsize\it shell}}>T_{A,\hbox{\scriptsize\it shell}}
   >T_{A,H}  .
\label{Carlipf6}
\end{align}
This will result in a net energy flux from shell $A$ into the black hole,
and from the black hole out to shell $B$.   It is not hard to check that
one can adjust the shell temperatures so that the flux into the black
hole balances the flux out.  The net result, then, is a flux of energy
from the ``cold'' shell $A$ to the ``hot'' shell $B$, violating the
second law of thermodynamics.

Arguments like this should be treated cautiously.  Historically, the
generalized second law of thermodynamics has shown remarkably
resilience, often involving subtle and unexpected effects.  There
may be natural Lorentz-violating theories in which all matter shares
a single universal horizon, for instance, or theories in which 
the black hole interior carries hidden
entropy.  But the apparent paradox shows, at least, that simple 
thermodynamic arguments about black holes may place surprising
restrictions on nongravitational physics.

\section{Approaches to black hole statistical mechanics \label{stata}}

In ordinary thermodynamic systems, thermodynamic properties
reflect the statistical mechanics of underlying microscopic states.
Entropy, in particular, is a measure of the number of accessible
states.  One might hope for the same to be true for black holes.  Then,
since the Bekenstein-Hawking entropy depends on both Newton's
constant $G$ and Planck's constant $\hbar$, it is plausible that 
black hole thermodynamics is telling us something about quantum
gravitational states.

Black hole thermodynamics may therefore present us with a rare
opportunity.  There is very little that we actually know about quantum
gravity, so any insight is likely to be valuable.  Like many opportunities,
though, this one is accompanied by a serious difficulty: in the absence
of a quantum theory of gravity, how do we start to understand the 
quantum states of a black hole?

Fortunately, while we do not yet have a complete quantum theory of
gravity, we have a number of research programs, at various levels
of maturity, attempting to develop such a theory.  Some of these
are advanced enough to be able to make predictions about at least
certain classes of black holes.  Indeed, as I shall discuss in section
9, the real problem may be not that we have no microscopic
explanation of black hole entropy, but that we have too many.

\subsection{``Phenomenology'' \label{phenoma}}

From the earliest days of black hole thermodynamics, there was an 
obvious simple ``phenomenological'' model for Bekenstein's area law 
\cite{Bekenstein}: one had to merely assume that black hole horizon 
area was quantized, with discrete ``plaquettes'' of area on the order 
of the Planck area.  The simplest choice --- partially justified by
the observation that the horizon area is an adiabatic invariant 
\cite{Barvinsky} --- would a discrete, evenly spaced area spectrum.  

Suppose, for instance, that the Bekenstein-Hawking entropy were 
\emph{exactly} the logarithm of the number of states.  This number
is, of course, an integer, so successive values of the area would have to 
differ by an amount
\begin{align}
\Delta A = 4\hbar G\ln k
\label{Carlipg0}
\end{align}
for some integer $k$.  As Hod first observed \cite{Hod}, one could then
appeal to the Bohr correspondence principle to determine the area
spacing.  While black holes have no stable excitations, they do have 
damped ``ringing modes,'' called quasinormal modes, with complex 
frequencies \cite{Siopsis}.  The most strongly damped quasinormal 
modes of the Schwarzschild black hole have frequencies
\begin{align}
\mathop{Re}\omega \sim \ln 3/8\pi GM   .
\label{Carlipg01}
\end{align}
The Bohr correspondence principle would then suggest that these
frequencies should correspond to transitions between adjacent area 
eigenstates,
%\footnote{The restriction to the most highly damped modes
%is justified by the argument that these short-lived modes are most 
%likely to correspond to single transitions between adjacent states.}
with energies
\begin{align}
\Delta M =  \hbar\omega \ \Rightarrow \
\Delta A/32\pi G^2 M = \hbar\ln 3/8\pi GM  ,
\label{Carlipg02}
\end{align}
which exactly matches (\ref{Carlipg0}) with $k=3$.

Arguments of this sort are suggestive, but must also be treated with
care.  For instance, within the same framework, Maggiore \cite{Magg}
has argued --- based in part on an analogy with a damped harmonic
oscillator --- that relevant frequency one should use for the Bohr
correspondence principle is not the real part $\mathop{Re}\omega$
of the quasinormal frequency, but the modulus $|\omega|$.  For
highly damped modes,
\begin{align}
|\omega| \sim  \frac{1}{4GM} \left(n+\frac{1}{2}\right)   ,
\label{Carlipg03}
\end{align}
and the Bohr correspondence principle then gives
\begin{align}
\Delta A = 8\pi\hbar G   .
\label{Carlipg04}
\end{align}
A wide range of simple models have been explored --- Medved's
review \cite{Medved} gives a partial list --- and they do not agree
about the area spectrum.  The lesson seems to be, not surprisingly, 
that thought experiments and analogies can only go so far; one needs
 a genuine quantum theory of gravity to obtain a real answer.

In  the remainder of this section, I will briefly summarize some of
the more specific approaches to black hole statistical mechanics from
the point of view of particular models of quantum gravity.

\subsection{Entanglement entropy}

Consider a space that is divided into two disjoint pieces  $A$
and $B$.  Suppose a quantum system on that space is 
characterized by a pure state $|\Psi\rangle$, with a corresponding
density matrix $\rho_{AB} = |\Psi\rangle\langle\Psi|$.  The von
Neumann entropy of such a pure state,
\begin{align}
S_{vN} = \mathop{Tr} \rho \ln\rho 
\label{Carlipg1}
\end{align}
is, of course, zero.

Now, however, imagine an observer who is restricted to the region $A$,
and cannot observe anything in $B$.  All observations of such an 
observer can be described by a reduced density matrix,
$\rho_A = \mathop{Tr}_B\rho_{AB}$, whose entropy will in general 
be nonzero.  For a local theory with a finite-dimensional Hilbert space,
for instance, we can write the state $|\Psi\rangle$ in a Schmidt
decomposition,
\begin{align}
|\Psi\rangle = \sum c_i |\phi_i\rangle|\psi_i\rangle  ,
\label{Carlipg2}
\end{align}
where $\{|\phi_i\rangle\}$ and $\{|\psi_i\rangle\}$ are orthonormal
bases of states in regions $A$ and $B$.  Then
\begin{align}
\rho_A = \sum |c_i|^2|\phi_i\rangle\langle\phi_i|
\label{Carlipg3}
\end{align}
and the von Neumann entropy of region $A$ is simply
\begin{align}
S_A = \mathop{Tr} \rho_A \ln\rho_A = \sum |c_i|^2\ln|c_i|^2 ,
\label{Carlipg4}
\end{align}
the Shannon entropy for the probabilities $p_i = |c_i|^2$.

This ``entanglement entropy'' is now commonplace in quantum theory,
but it was first introduced by Sorkin in the context of black hole
thermodynamics \cite{Sorkin}.  The idea is that an observer outside
the horizon (``region $A$'') knows nothing about the state inside
the horizon (``region $B$''), and therefore sees a mixed state with
nonvanishing entropy.  This picture arises naturally from the eternal
black hole of section \ref{eterna}; the 
density matrix that gives the entropy (\ref{Carlipg4}) is the same as
the thermal density matrix that gives the Hawking spectrum.   For
many states. including the vacuum state, the entanglement entropy
for a black hole is proportional to the horizon area \cite{Sorkin,Bomb,Sred,FroNov},
since the main contribution comes from correlations among degrees
of freedom very close to the horizon, for which the ``bulk'' state
is irrelevant.

The \emph{coefficient} of this area term, on the other hand, diverges, 
for  essentially the same reason: the more closely degrees of 
freedom are localized to the horizon, the higher their energies.
The simplest cut-off, a ``brick wall'' approximately one Planck
length in proper distance from the horizon \cite{tHooft}, gives a
coefficient for the Bekenstein-Hawking entropy of the right order
of magnitude.  In general, though, the exact value depends
sensitively on the number and type of entangled fields --- that
is, on the particle content of the Universe.

A possible solution to this ``species problem'' comes from the
observation by Susskind and Uglum that the same modes responsible
for the entanglement entropy also renormalize Newton's constant,
which also appears in the Bekenstein-Hawking entropy formula
\cite{SussUg}.  It is plausible that these contributions cancel.  In
particular, Cooperman and Luty \cite{Coop} have recently argued 
that for a certain class of states defined by a path integral, the
leading renormalized term in the entanglement entropy is given by
the corresponding renormalized Bekenstein-Hawking formula,
independent of the particle content of the theory.

Inspired by the AdS/CFT correspondence of string theory, Ryu
and Takayanagi have proposed an alternative ``holographic'' 
approach to defining and regulating entanglement entropy
\cite{Ryu,Hubeny}.  In the description I have given above, the
regions $A$ and $B$ lie on a $(d-1)$-dimensional spacelike 
slice of a $d$-dimensional spacetime.  Ryu and Takayanagi
propose embedding this spacetime as the asymptotic boundary 
of $(d+1)$-dimensional anti-de Sitter space.  The boundary 
between regions $A$ and $B$ can then be extended into the ``bulk'' 
anti-de Sitter space as a $d$-dimensional minimal surface; Ryu 
and Takayanagi propose that the entanglement entropy should 
be proportional to the area of this surface.  While this approach 
is obviously inspired by black hole thermodynamics, can be 
proven without any assumptions about black holes, at least for 
the static case \cite{Lewkowycz}.  The holographic entanglement 
entropy can then be used to determine the entropy of a black hole, 
and one again obtains the Bekenstein-Hawking expression, with no 
divergences or species problem \cite{Emparanb}.

\subsection{String theory}

The leading research program in quantum gravity today is string 
theory.  I will not attempt to summarize the theory here,
but will instead focus on three rather
different approaches to black hole statistical mechanics within
string theory.

\subsubsection{Weakly coupled strings and branes \label{weaka}}

The first successful calculation of black hole entropy in string theory,
by Strominger and Vafa \cite{StromVafa}, looked at a class of  multiply
charged, extremal (that is, maximally charged), supersymmetric (BPS)
black holes in five dimensions.  As classical objects, such black holes 
are completely determined by their charges: the mass, horizon area, 
and all other macroscopic quantities can be expressed as functions 
of the charges.  As string theoretical objects, on the other hand,
they are strongly bound states of the fundamental constituents of
string theory, strings and D-branes.  Strominger and Vafa suggested
``turning down'' the strength of the gravitational interaction, until
a black hole became a weakly coupled system of strings and 
D-branes.  These constituents would still have the prescribed charges, 
but at weak coupling one could also count the number of states with such
charges.  The result, translated back into the black hole parameters, 
exactly matched the Bekenstein-Hawking entropy.  

These results were quickly extended to a large number of extremal
and near-extremal black holes, and, through dualities, to some
particular nonextremal black holes as well \cite{Peet,Das}.
A straightforward extension of the argument gives Hawking radiation,
coming from decay of excited states of the constituent strings and 
D-branes, complete with the correct greybody factors \cite{Mal}.
One might worry about the consistency of the procedure: the 
process of turning down the gravitational coupling need not preserve 
the number of states.  For supersymmetric black holes, though, the
number of states is protected by nonrenormalization theorems.  
For nearly supersymmetric black holes, the procedure continues to
work well, perhaps even better than one might expect; far from 
extremality, it becomes harder to obtain the right factor of $1/4$ 
in the entropy.

There is one peculiar feature of these calculations, which I will return to
in section \ref{universa}.  While the final result is an expression for 
entropy in terms of horizon area, the connection is indirect.  On the
strong coupling side, one may start with the horizon area, but one
must reexpress it in terms of charges.  On the weak coupling side,
one may start with the number of states, but one must again reexpress
it in terms of charges.  It is only when the two processes are compared
that one obtains the Bekenstein-Hawking entropy.  This means that
each new type of black hole requires a new calculation; some crucial
aspect of universality is missing.

\subsubsection{Fuzzballs \label{fuzza}}

Mathur has proposed running the analysis of the 
preceding section backwards \cite{Mathur}.  Suppose one starts on
the weak coupling side, with a particular collection of strings and
D-branes whose charges match those of a desired black hole.  One
may then turn the gravitational coupling up, and try to see what 
geometry appears at strong coupling.  The result is typically 
\emph{not} a black hole, but rather a ``fuzzball,'' a configuration 
with no horizon and no singularity, but with a geometry that looks 
very much like that of a  black hole outside a would-be horizon 
\cite{Mathurb,Mathurc}.  This phenomenon seems to require 
the extra dimensions available in string theory, but it provides
an intriguing new picture of black holes.  For example, Hawking 
radiation can appear as a result of a classical instability in a
fuzzball geometry \cite{Chowdhury}.

 In a few special cases, one can count the number of such 
classical ``fuzzball'' geometries and reproduce the 
Bekenstein-Hawking entropy.  In general, though, it is
likely that many of the states relevant for determining the
Bekenstein-Hawking entropy will not have classical geometric
descriptions.

\subsubsection{The AdS/CFT correspondence \label{adscfta}}

A third approach approach to black 
holes in string theory exploits Maldacena's celebrated AdS/CFT 
correspondence \cite{Malda,AGMOO}.  This extremely well-supported 
conjecture states that string theory in a $d$-dimensional 
asymptotically anti-de Sitter spacetime is dual to a conformal field 
theory in a flat $(d-1)$-dimensional space that can, in a sense, 
be viewed as the boundary of the AdS spacetime.  For asymptotically 
anti-de Sitter black holes, this means that one can, in principle,
compute the entropy by counting states in a (nongravitational) dual 
conformal field theory.  

The most straightforward application of this duality occurs for the
(2+1)-dimensional BTZ black hole discussed in section \ref{BTZa}.
Here, the dual description is given by a two-dimensional  conformal 
field theory.  As I shall describe in more detail in section \ref{universa},
two-dimensional conformal field theories have an exceptionally
large symmetry group that controls many of their properties.  In
particular, the density of states is determined by a single parameter
$c$, the central charge, which characterizes the conformal anomaly.
As we saw in section \ref{anoma}, anomalies can provide very useful
tools; here, they are powerful enough to determine the density of
states.
 
For asymptotically anti-de Sitter gravity in 2+1 dimensions, the 
central charge is dominated by a classical contribution, which was 
discovered some time ago by Brown and Henneaux \cite{BrownHen}.  
Strominger \cite{Strominger} and Birmingham, Sachs, and Sen 
\cite{BSS} independently realized that this result could be used to 
compute the BTZ  entropy, precisely reproducing the 
Bekenstein-Hawking expression.  As noted in section \ref{BTZa},
the same methods can be used to compute Hawking radiation for
the BTZ black hole.

While this result appears to be rather specialized, it has an important
extension.  Many higher dimensional near-extremal black holes,
including black holes that are not themselves asymptotically anti-de 
Sitter, have a near-horizon geometry of the form 
$\mathit{BTZ}\times\mathit{trivial}$, where the ``trivial'' part merely 
renormalizes constants in the calculation of entropy.  As a result, the 
BTZ results can be used to obtain the entropy of a large class of string
theoretical black holes, including most of the black holes whose states 
could be counted in the weak coupling approach of section \ref{weaka} 
\cite{Skenderisb}.

\subsection{Loop quantum gravity \label{loopa}}

The second major research program in quantum gravity is loop 
quantum gravity, or ``quantum geometry.''  I will again focus 
only on those elements relevant to black hole thermodynamics.

The key features  relevant to this question are these:
\begin{enumerate} 
\item A basis for the kinematical states of the theory is given by
spin networks, graphs with edges labeled by 
spins and vertices labeled by $\mathrm{SU}(2)$ intertwiners.
\item Given any surface $\Sigma$, each spin network is an eigenstate 
of the area operator ${\hat A}_\Sigma$, with eigenvalue
\begin{align}
A_\Sigma = 8\pi\gamma G \sum_j \sqrt{j(j+1)} 
\label{Carlipg5}
\end{align}
where the sum is over the spins $j$ of edges of the spin network 
that cross $\Sigma$.  Note that this area is quantized, but in a
rather complicated way, with a spectrum that can be shown to have
an elaborate substructure \cite{Corichi,Agullo}.
\item Operators such as the area and volume  depend on a
parameter $\gamma$, the Barbero-Immirzi parameter.
The significance of this 
parameter is quite poorly understood; theories with different values of 
$\gamma$ are probably inequivalent, but it has been suggested that 
$\gamma$ may not appear in properly renormalized observables 
\cite{Jacobsone} or in a somewhat different approach to quantization 
\cite{Alexandrov}.
\item The physical states are obtained from the spin networks by
imposing the Hamiltonian constraint, a procedure that is not yet
completely under control.
\end{enumerate}

\subsubsection{Microcanonical approach}

Given this structure, a natural first step is to choose the surface
$\Sigma$ to be a black hole horizon and count the number of spin 
network states that give a prescribed area \cite{Krasnov,Rovellib}.   
This is a bit tricky, since the true event horizon  is 
defined by the global properties of the spacetime.  On the other hand, 
we know that Hawking radiation depends only on local
properties, so a local characterization
should be adequate.  The relevant definition is probably that of an
``isolated horizon,'' a null surface with vanishing expansion that
obeys the laws of black hole mechanics \cite{Ashisol,Ashisolb}.

As we shall see in section \ref{universa}, the specification
of such a horizon is a sort of boundary condition, which requires 
the introduction of boundary terms in the Einstein-Hilbert action.
For loop quantum gravity, these boundary terms induce a three-%
dimensional Chern-Simons action on the horizon.  The careful 
version of the state-counting \cite{ABCK,ABK} can then be
translated into a counting of Chern-Simons states, a well-understood
procedure \cite{Witten_Jones}, although with slightly subtle 
combinatorics.  The ultimate result is a microcanonical  black hole 
entropy \cite{Domagala,Meissner}
\begin{align}
S = \frac{\gamma_M}{\gamma}
  \frac{\ A_{\mathrm{\scriptstyle hor}}}{4\hbar G} ,
\label{Carlipg6}
\end{align}
where $\gamma$ is the Barbero-Immirzi parameter and
\begin{align}
\gamma_M \approx .23753
\label{Carlipg7}
\end{align}
is a constant determined as the solution of a particular
combinatoric problem.  If one chooses $\gamma=\gamma_M$, 
one recovers the standard Bekenstein-Hawking entropy.  

The significance of this peculiar value $\gamma_M$ is not 
understood, and it may reflect an inadequacy in the quantization 
procedure or the definition of the area operator \cite{Alexandrov}.  
Note, though, that the Barbero-Immirzi parameter appears only
in the combination $G\gamma$, where $G$ is the ``bare''
Newton's constant, \emph{i.e.}, the parameter appearing in the action.
The relevant constant in the Bekenstein-Hawking entropy, however,
is the renormalized value \cite{Jacobsone}, and since it is not known
how Newton's constant is renormalized in loop quantum gravity,
the ``physical'' value of $\gamma$ remains uncertain.
Ideally, one could address this question by computing the attraction
between two test masses in the Newtonian limit, but this seemingly
straightforward problem is not yet solved --- loop quantum gravity
is defined at the Planck scale, and the extrapolation to
physically realistic distances remains extremely difficult.

In any case, though, once $\gamma$ is fixed for one type of black
hole --- the Schwarzschild solution, for instance --- the loop 
quantum gravity computations give the correct entropy for a wide 
variety of others, including charged black holes, rotating black holes, 
black holes with dilaton couplings, black holes with higher genus 
horizons, and black holes with arbitrarily distorted horizons 
\cite{AshtekarLew,Engle}.  In particular, there is no need 
to restrict oneself to near-extremal black holes.  

As an interesting variation within this framework \cite{Livine}, 
one may again look at the horizon area (\ref{Carlipg5}), but instead 
of counting states of the boundary Chern-Simons theory one may 
count the number of ways the boundary spins can 
be joined to a single interior vertex.  This means, in essence, that 
one is completely course-graining the interior state of the black
hole, much as one does in an entanglement entropy computation; 
there are, in fact, interesting relationships to entanglement
entropy.  The end result is again an entropy proportional to 
the horizon area, but now requiring a different value of the 
Barbero-Immirzi parameter to match the Bekenstein-Hawking 
result.

\subsubsection{Other ensembles}

The microcanonical approach described above involves a hidden
assumption: that all spin networks that give the same horizon area
occur with equal probability.  This is a standard assumption in
ordinary thermodynamics, but gravitating systems are not quite
ordinary.  Moreover, the microcanonical approach uses only
properties of the ``kinematical Hilbert space,''  that is, the space
of states before the Hamiltonian constraint has been imposed.
But we know in other contexts that although Hawking radiation is
basically kinematical, the Bekenstein-Hawking entropy depends
on the dynamics \cite{Visser_essential}, including the Hamiltonian 
constraint.

In the past several years, interesting progress has been made in
moving away from thesse assumptions.  In particular,

-- The quantum Hamiltonian that generates translations along the
horizon has been identified \cite{Bianchi}, and yields 
a local temperature and energy, as measured by an idealized
particle detector, that agree with semiclassical expectations \cite{Frodden}.  
This is not yet an enumeration of microscopic states, but it is a version
of the thermodynamic computation of section \ref{detecta} in a
fully quantum gravitational setting.

-- One can define a sort of grand canonical ensemble, in which
the number of punctures --- that is, of edges of the spin network 
that cross the horizon --- appears with a corresponding chemical 
potential \cite{Ghosh}.  The Barbero-Immirzi parameter still appears, 
but in a different form: the entropy becomes
\begin{align}
S =  \frac{\ A_{\mathrm{\scriptstyle hor}}}{4\hbar G}  + N\sigma(\gamma)
\label{Carlipg8}
\end{align}
where $\sigma(\gamma)$ is  a Lagrange multiplier that vanishes when
$\gamma$ takes the value (\ref{Carlipg7}).  It has been argued that if
one considers additional couplings to matter, and makes the
``holographic'' assumption that the matter entropy near the horizon 
scales with the area, then the Lagrange multiplier term vanishes in
the classical limit, giving back the standard Bekenstein-Hawking 
entropy \cite{Perez}.

-- The Barbero-Immirzi parameter can in principle take any value,
but there is one ``natural'' value, $\gamma=i$.  At this value, the
Ashtekar-Sen connection of loop quantum gravity is self-dual,
and the Hamiltonian constraint becomes much simpler.  Loop quantum
gravity is unfortunately much harder to work with when $\gamma=i$,
because of the need for added constraints to ensure that the metric
is real, but several recent computations \cite{Geiller,Achour,Carlip_loop}
indicate that this choice also gives the correct Bekenstein-Hawking
entropy, perhaps even more simply.

\subsection{Induced gravity}

Induced gravity, as first proposed by Sakharov \cite{Sakharov},
is a model in which the Einstein-Hilbert action is not fundamental,
but appears a consequence of the presence of other fields.  If one starts with 
a theory of quantum fields propagating in an initially nondynamical
curved spacetime, counterterms from renormalization will automatically 
induce a gravitational action, which almost always includes an 
Einstein-Hilbert term at lowest order \cite{Adler}.  In Sakharov's
language, the resulting spacetime dynamics is a kind of 
``metric elasticity'' induced by quantum fluctuations of matter.

One can write down an explicit set of ``heavy'' fields that 
can be integrated out in the path integral to induce the Einstein-Hilbert 
action.  The gravitational counterterms normally have divergent
coefficients that must be renormalized, but by including an
appropriately chosen collection of nonminimally coupled scalar fields, 
one can obtain finite values for Newton's constant and the 
cosmological constant \cite{Frolovx}.  

Given such a model, one can then count quantum states of the 
``heavy'' fields in a black hole background \cite{Frolovx,Frolovb}.
Some subtleties occur in the definition of entropy in the
presence of nonminimally coupled fields, but in the end the 
computation reproduces the standard Bekenstein-Hawking 
expression, with the correct coefficient.  Furthermore, 
if one dimensionally reduces to a two-dimensional conformally 
invariant system near the horizon,  
one can count states by standard methods of conformal field
theory \cite{Frolovc}, as described below in section \ref{universa}.  
This offers a new, and apparently completely different, picture 
of black hole microstates as those of the ordinary quantum fields 
responsible for inducing the gravitational action.

\subsection{Other approaches}

Just as in black hole thermodynamics, there are a number of
other approaches to black hole statistical mechanics that also
show promise.  For example:

\begin{itemize}
\item Classical black holes do not have excited states, but they
have characteristic damped oscillations, called quasinormal 
modes \cite{Siopsis}.  York has argued that these modes should be 
quantized \cite{York}, and as noted in section \ref{otherb}, obtains a 
value close to the Bekenstein-Hawking entropy with an approximate
quantization.

\item The ``causal set'' program is an attempt to quantize gravity
that starts by replacing a continuous spacetime by a discrete set
of points with specified causal relations \cite{Sorkin_caus}.  While
it is not yet known how to perform an exact calculation of black
hole entropy, there
are indications that one can obtain a reasonable value by counting
the number of points in the future domain of dependence of a
cross-section of the horizon \cite{Rideout} or the number of
causal links crossing the horizon \cite{Dou}.

\item Zurek and Thorne have shown that there is sense in which 
the entropy of a black hole counts the number of distinct ways
in which a black hole with prescribed macroscopic properties
such as mass and angular momentum can be formed from the
collapse of quantum matter \cite{Zurek}.
\end{itemize}

\subsection{Logarithmic corrections}

While many approaches to quantum gravity give the correct Bekenstein-Hawking
entropy, they may differ on the next order corrections.  In general, one expects
\begin{align} 
S_{\scriptscriptstyle\mathrm{BH}}
 = \frac{\ A_{\mathrm{\scriptstyle hor}}}{4\hbar G}  
+ \alpha \ln\left(\frac{\ A_{\mathrm{\scriptstyle hor}}}{4\hbar G} \right)
+ \dots
\label{Carlipx1}
\end{align}
where the coefficient $\alpha$ may depend on the quantum theory.  There
are some arguments for a ``universal'' answer that, as in section \ref{universa},
depends on conformal field theory \cite{Carliplog}, but these are not
conclusive, and particular models seem to give varying results 
\cite{Ghoshb,Medvedx,Sen,ENPP}.  The problem is further complicated
by the fact that the logarithmic terms in (\ref{Carlipx1}) may differ
depending on one's choice of thermodynamic ensemble \cite{Gour}.

\section{The holographic conjecture}

In ordinary thermodynamic systems, entropy is an extensive quantity, 
scaling with volume.  For black holes, in contrast, the entropy scales
with area.  It is not so surprising that entropy is not quite extensive
for a gravitating system: gravity cannot be shielded, so when
one increases the size of a system one is necessarily changing the
internal interactions.  Nor is it surprising that entropy cannot be
simply defined in terms of a density that can be integrated over
a volume: there is a general result that, as a consequence of
diffeomorphism invariance, observables in general relativity cannot 
be defined in terms of local densities \cite{Torre}.  Still, an 
entropy proportional to area is a rather dramatic departure.

't Hooft \cite{Hooft_holo} and Susskind \cite{Suss_holo} have
proposed that this feature is not unique to black holes, but is
a general property of any gravitating system.  They
suggest that the entropy inside \emph{any} region, whether
or not it is a black hole, should scale as the area rather than
the volume.  This conjectured ``holographic principle'' would 
imply that our usual view of local physics drastically overcounts 
degrees of freedom.

As a first argument for the plausibility of this holographic viewpoint
\cite{Yurtsever}, consider an approximately spherical surface 
$\mathcal{S}$ with a surface area $A_{\mathcal{S}} = 4\pi R^2$.  
Let us try to fill the interior of this surface with quantum excitations.  
To be contained in the region, each excitation should have a wavelength 
no larger than $2R$, and thus an energy $E\gtrsim \hbar/2R$.  $N$ such
excitations will have an energy $E\gtrsim \hbar N/2R$, and to avoid 
forming a black hole, we must have $R\gtrsim 2GE$.  Hence
\begin{align}
N \lesssim R^2/\hbar G
\label{Carliph1}
\end{align}
If we interpret $N$ as the number of degrees of freedom in the 
system, we see that (\ref{Carliph1}) implies a holographic bound.

As a second argument, suppose the surface $\mathcal{S}$ now
contains \emph{almost} enough matter to form a black hole.
Now surround the region by a collapsing shell of matter, with 
enough additional mass to form a black hole.  The initial state has 
an entropy
$$S_{\scriptscriptstyle\mathrm{init}} = S_{\scriptscriptstyle\mathrm{interior}}
  + S_{\scriptscriptstyle\mathrm{shell}}$$
while the final state has an entropy
$$S_{\scriptscriptstyle\mathrm{final}} = \frac{A_{\mathcal S}}{4\hbar G}$$
By the generalized second law of thermodynamics, 
$S_{\scriptscriptstyle\mathrm{final}}
\ge S_{\scriptscriptstyle\mathrm{init}}$.  Hence the matter initially
inside $\mathcal{S}$ must have had no more entropy than 
that of a black hole.

Neither of these arguments is conclusive.  The first implicitly
assumed that the region inside $\mathcal{S}$ was flat; one may be 
able to fit more degrees of freedom into a curved space. There are, in fact, 
classical configurations --- so-called ``monsters'' 
--- in which the entropy within a region is greater than that of the 
corresponding black hole \cite{Hsu}, although all known examples 
form from initial singularities (they cannot be ``made in the laboratory'') 
and collapse quickly to form black holes.   

The second argument fails for a more subtle reason: the area of
$\mathcal{S}$ is not a gauge invariant quantity \cite{Flanagan,Bousso}.  
We are really imagining a timelike ``world tube'' traced 
out by $\mathcal{S}$ and asking its cross-sectional area at a fixed 
time.  But this area depends on the choice of time-slicing, and by 
choosing a slice that ``wiggles'' enough in a timelike direction --- 
and is therefore nearly null over much of its intersection with the 
world tube --- we can make the area as small as we like.

\begin{figure}
\centerline{\includegraphics[height=1.8in]{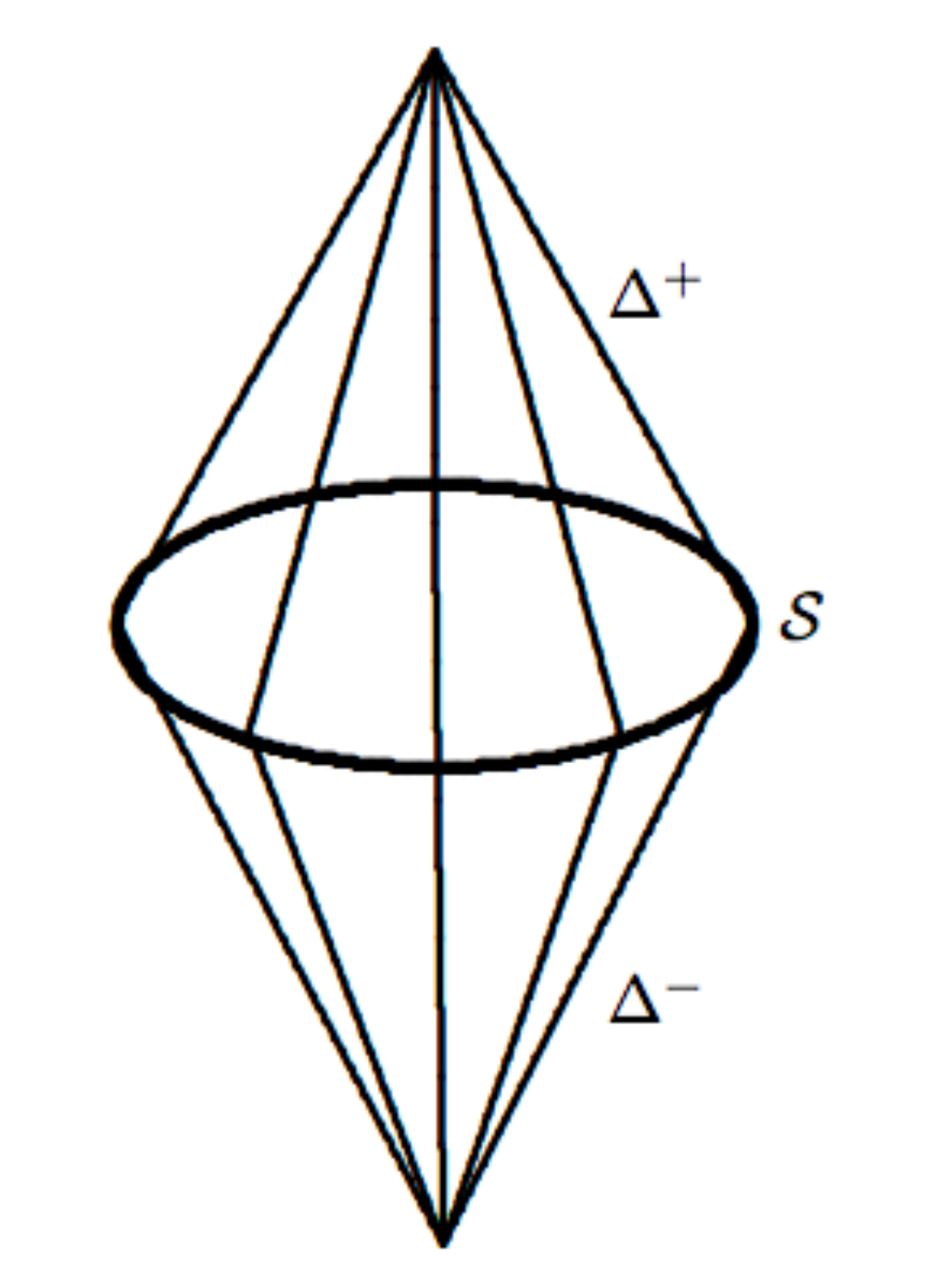}}
\caption{Future-directed null geodesics orthogonal to
$\mathcal{S}$ form two light sheets, usually one expanding outward 
(not shown) and one collapsing inward ($\Delta^+$).  The same holds
for past-directed null geodesics.  In this figure, symmetry causes 
the light sheets to collapse to single points.  In general, they will 
end at caustics where neighboring geodesics cross 
and the expansion becomes positive.}
\label{Carlipfig4}
\end{figure}
One might worry that the same problem occurs for black holes, and that
the ``horizon area'' might also not be well-defined.  In fact, black holes
evade the problem, essentially because the horizon is not timelike, but
null (and, for stationary black holes, nonexpanding).  Bousso \cite{Boussob}
has proposed a covariant entropy bound that exploits a similar idea.
He starts with a closed surface $\mathcal{S}$, no longer necessarily
spherical, and extends it not as a timelike world tube, but
as a ``collapsing'' light sheet, a null hypersurface of decreasing area
extending to the past and future of $\mathcal{S}$, as illustrated in figure
\ref{Carlipfig4}.  Bousso's conjecture is that the total 
entropy flux through either light sheet --- $\Delta^+$ or $\Delta^-$
in the figure --- is bounded by $A_{\mathcal{S}}/4\hbar G$.  For the
special case of a black hole, $\Delta^-$ may be interpreted as the
event horizon, and the Bousso entropy bound is basically the generalized
second law of thermodynamics.

Classically, Bousso's conjecture can be proven correct, given suitable
energy conditions and bounds on the entropy flux 
and its gradient \cite{Flanagan,BoussoFlan}.  Quantum mechanically, 
the situation is less clear, though there is strong evidence for the case 
of free fields with negligible gravitational backreaction \cite{BoussoMal}.

Although the holographic principle in its larger sense remains a conjecture,
there is one case in which it has been dramatically successful.  The
AdS/CFT correspondence of string theory \cite{Malda,AGMOO} describes
the ``bulk'' physics of an asymptotically anti-de Sitter space in terms
of a lower-dimensional ``surface'' conformal field theory.  If this
holographic viewpoint can be extended beyond the anti-de Sitter case,
black hole thermodynamics will have had a fundamental impact on a far
wider field, profoundly altering our approach to fundamental physics.

\section{The problem of universality \label{universa}}

For a physicist working on quantum gravity,  black hole 
thermodynamics is, at first sight, a huge source of hope.
The Hawking temperature and Bekenstein-Hawking entropy are
``quantum gravitational,'' depending on both
Planck's and Newton's constants.  Moreover, as Wheeler's
famous aphorism states, ``a black hole has no hair'' ---  a classical
black hole has no  distinguishing characteristics 
beyond its mass, charges, and spins.  Hence if the entropy of a black 
hole is an ordinary statistical mechanical entropy, it seems that the 
only states it can be counting are quantum gravitational states.  We 
may thus have a window into quantum gravity.

Upon closer examination, though, the situation is not so simple.  As
we saw in section \ref{stata}, we suffer an embarrassment of riches: 
there seem to be many different quantum descriptions of black
hole statistical mechanics, with very different quantum
states, all giving the same entropy.  Moreover, while the simple
Bekenstein area law seems natural in some of these
approaches, in others it seems almost miraculous.

The ``problem of universality'' is really two problems, which are 
logically distinct, although their solutions may be related.  The
first is that black hole entropy has such a simple, universal value,
one-fourth of the horizon area in Planck units, regardless 
of the mass, spin, charges, or the number of dimensions.  Even
the horizon topology is irrelevant: the same area law holds for
black holes, black strings, black rings, and black branes.  The
second is that so many different approaches to quantum gravity
give the same result, regardless of any of the details of the
black hole states.

It is tempting to address the first problem with a claim 
that the horizon area must be quantized, as in section
\ref{phenoma}.  But such an answer is at best incomplete: it
does not explain the universal coefficient of $1/4$, nor does
it tell us why no other features matter.  There is an 
interesting explanation of the factor of $1/4$ in terms of the 
topology of the Euclidean black hole \cite{BTZa}, but it is not 
clear why this feature would be reflected in the counting of states.

A cynical answer to the second problem is that only quantum
theories that give the ``right'' result are published.
But this, too, is at best incomplete: it does not explain why any
such models work at all.  Consider, for example, the weakly 
coupled string approach of section \ref{weaka}.  As we saw
there, the theory gives the correct entropy for a large class of
near-extremal black holes; but each new type of black hole
requires a separate computation, and the relationship between
entropy and area only appears at the very last step.  Such  
miracles cry out for explanation.

In the remainder of this section, I will describe one attempt to
explain these miracles.  This is still very much a case of work in
progress, but there have been some hopeful signs.

\subsection{State-counting in conformal field theory}

Statistical mechanical entropy is basically a measure of the number
of quantum states, so we are looking for a universal mechanism to
explain the density of states of a black hole.  There is one context
in which such a mechanism is known: that of two-dimensional
conformal field theory.

The metric for a two-dimensional manifold can always be written 
locally as
\begin{align}
ds^2 = 2g_{z{\bar z}}dzd{\bar z}
\label{Carlipi1}
\end{align}
in terms of complex coordinates $z$ and $\bar z$.  Holomorphic 
and antiholomorphic coordinate changes $z\rightarrow z + \xi(z)$,
${\bar z}\rightarrow {\bar z} + {\bar\xi}({\bar z})$ merely rescale the 
metric, and  provide the basic symmetries of a conformal field theory.   
Unlike the conformal symmetry group in higher dimensions, the
two-dimensional group has infinitely many generators, denoted 
$L[\xi]$ and ${\bar L}[{\bar\xi}]$.  These satisfy a Virasoro algebra 
\cite{FMS},
\begin{align}
&[L[\xi],L[\eta]] 
   = L[\eta\xi' - \xi\eta']   + \frac{c}{48\pi}\int dz\left( \eta'\xi'' - \xi'\eta''\right) \nonumber\\
&[{\bar L}[{\bar\xi}],{\bar L}[{\bar\eta}]] 
   = {\bar L}[{\bar\eta}{\bar\xi}' - {\bar\xi}{\bar\eta}']  + \frac{{\bar c}}{48\pi}%
   \int d{\bar z}\left( {\bar\eta}'{\bar\xi}'' - {\bar\xi}'{\bar\eta}''\right) \nonumber\\
&[ L[\xi],{\bar L}[{\bar\eta}]] = 0 , 
\label{Carlipi2}
\end{align}
uniquely determined by the values of the two constants $c$ and $\bar c$, the 
central charges.  As in ordinary field theory, the zero modes $L_0$ and ${\bar L}_0$ 
of the symmetry generators are conserved quantities, the ``conformal weights,''
which can be seen as linear combinations of mass and angular momentum.

Two-dimensional conformal symmetry is remarkably powerful.  In particular,
with a few mild restrictions, Cardy has shown that the asymptotic density of
states of any two-dimensional conformal field theory is given by \cite{Cardy,Cardy2}
\begin{alignat}{5}
&\ln\rho(L_0) \sim 2\pi\sqrt{\frac{cL_0}{6}}, \qquad 
&&\ln{\bar\rho}({\bar L}_0) \sim 2\pi\sqrt{\frac{{\bar c}{\bar L}_0}{6}} \qquad
&&\hbox{(microcanonical)} \nonumber\\
&\ln\rho(T) \sim \frac{\pi^2}{3}cT, \qquad 
&&\ln{\bar\rho}(T) \sim \frac{\pi^2}{3}{\bar c}T
&&\hbox{(canonical)}
\label{Carlipi3}
\end{alignat}
where $T$ is the temperature.  The entropy is thus uniquely determined by a 
few parameters, regardless of the details of the conformal field theory --- just 
the kind of universal behavior we would like for black holes.

\subsection{Application to black holes}

Black holes are neither conformally invariant nor two-dimensional, so one
might wonder whether the preceding results are relevant. There is a sense,
however in which the near-horizon region is \emph{nearly}  conformally
 invariant: the extreme red shift washes out dimensionful quantities such
as masses for an observer who remains outside the black hole \cite{Camblong}.
The same red shift also makes transverse excitations negligible relative to 
those in the $r-t$ plane, effectively reducing the problem to two dimensions; 
this is the same dimensional reduction that allowed many of the anomaly-based
calculations described in section \ref{anoma}.

Conformal techniques of this type were first applied to the (2+1)-dimensional
BTZ black hole \cite{Strominger,BSS}.  Here, the connection to two-dimensional
conformal field theory is clear: the asymptotic boundary at infinity is a 
two-dimensional cylinder, and the canonical generators of diffeomorphisms
obey a Virasoro algebra.  Moreover, as Brown and Henneaux showed long
ago \cite{BrownHen}, this Virasoro algebra has a classical central charge,
whose presence can be traced back to the need to add boundary terms to
the canonical generators.  The conformal weights $L_0$ and ${\bar L}_0$ 
for the BTZ black hole have simple expressions in terms of charge and 
mass, and the microcanonical form of the Cardy formula (\ref{Carlipi3}) 
then gives the standard Bekenstein-Hawking entropy.

The generalization to higher dimensional black holes \cite{Carlipcfta,%
Carlipcftb,Carlipcftc} is more subtle, and not so firmly established, but there 
has been some significant progress.\footnote{See my review \cite{Carlipcftd}
for a more thorough treatment.}  The key trick is to treat the horizon as 
a sort of boundary --- not in the sense that matter cannot pass through
it, but in the sense that it is a place where one must impose ``boundary
conditions,'' namely the condition that it is a horizon.  As in the BTZ
case, the canonical generators of diffeomorphisms acquire boundary
terms, and for a number of sensible ``stretched horizon'' boundary
conditions these pick out a Virasoro subalgebra with a calculable central
charge.  Similar techniques have been used for the near-horizon geometry
of the near-extremal Kerr black hole \cite{GHSS}, again giving the expected
Bekenstein-Hawking entropy.

\subsection{Effective descriptions \label{effecta}}

While conformal methods may allow a ``universal'' computation of
Bekenstein-Hawking entropy, they do not tell us a lot about the quantum
states of the black hole.  In one sense, this is a good thing: the point,
after all, is to explain why many different descriptions of the states
yield the same entropy.  Still, one may be able to learn something
useful about ``effective'' descriptions.

It is well known that the presence of a boundary, either at infinity or 
at a finite location, alters the canonical generators of diffeomorphisms 
by requiring the addition of boundary terms \cite{Regge}.   The 
exact form of these boundary terms depend on the choice of boundary 
conditions, and the resulting generators necessarily respect these 
boundary conditions.  As a result, some of the ``gauge transformations''
of the theory without boundary --- those that change the boundary
conditions --- are no longer invariances of the theory.  This boundary
symmetry-breaking increases the number of states in the theory: 
states that would previously have been considered gauge-equivalent
are now distinct \cite{Carlip_wouldbe}.  Moreover, these new states 
appear only at the boundary, since the gauge symmetry remains
unbroken elsewhere.

This phenomenon is not quite the same as the Goldstone mechanism
\cite{Goldstone}, but it is similar in spirit.  By analogy, it suggests a 
possible effective description of black hole horizon states in terms of 
the parameters that label the boundary-condition-breaking ``would-be
diffeomorphisms.''  For the case of the BTZ black hole, such a
description can be made explicitly \cite{CarlipBTZb,Carlipasymp}.  
The effective field theory is a particular conformal field theory, Liouville 
theory, with a central charge that matches the Brown-Henneaux value.  
The counting of states in this theory is poorly understood: technically,
the relevant states are in the ``nonnormalizable sector,'' which is not
under good control.  But as discussed in section \ref{BTZa}, one can 
couple the Liouville theory to external matter and recover a correct
description of Hawking radiation.

The effective description described here is holographic, and
is reminiscent of the ``membrane paradigm'' for black holes 
\cite{membrane,Parikhz}, in which a black hole is also described by 
a collection of degrees of freedom at the horizon.  It is possible
that a better understanding of these degrees of freedom could
help us understand the information loss problem, the next
topic of this review.

\section{The information loss problem}

Let us turn finally to one of the most puzzling aspects of black hole
thermodynamics \cite{Hawking_info}, the ``information loss paradox.''  
This is a topic that is very much in flux, with nothing near a consensus 
concerning its resolution.  The purpose of this section is therefore 
simply to introduce some of the central ideas.

Consider a process in which a shell of quantum matter in a pure state 
is allowed to collapse to form a black hole, which then evaporates
completely into Hawking radiation.  It appears that this process has
allowed an initial pure state to evolve into a final highly mixed
(thermal) state.  But this is not possible in quantum mechanics:
unitarity requires that a pure state evolve to a pure state.  If one
thinks of the von Neumann entropy as a measure of lack of information,
this process has led to an unallowable loss of information.

This argument contains several hidden assumptions, of course, any
one of which could point to a resolution.  We have assumed that a
pure state \emph{can} collapse to form a black hole; that the final
state consists only of Hawking radiation, with no long-lived  
``remnants'' of the black hole; that the final spacetime is simply
nearly flat space, with no residue of the singularity and no
``baby universes'' that could swallow up lost correlations; and
that the Hawking radiation is genuinely thermal, with no hidden
correlations.  Let us consider each of these in turn.

\subsection{Nonunitary evolution}

\begin{figure}
\centerline{\includegraphics[height=1.8in]{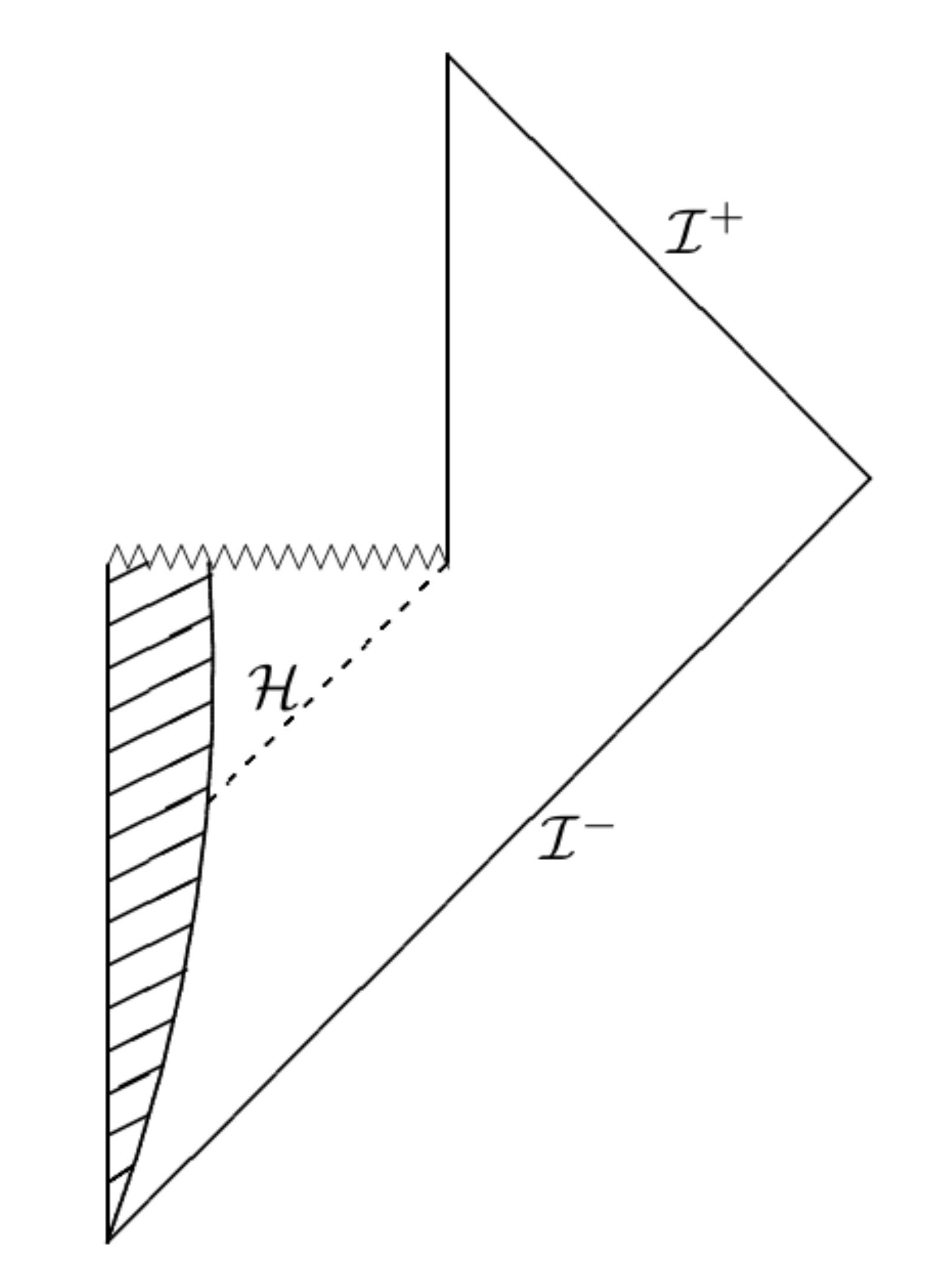}}
\caption{A possible Carter-Penrose diagram for an evaporating black
hole}
\label{Carlipfig5}
\end{figure}

One simple answer to the information loss problem is simply that 
quantum mechanics is not quite correct, and the evolution is not
unitary.  Figure \ref{Carlipfig5} shows a proposed Carter-Penrose
diagram for an evaporating black hole.  In
this picture, it is clear where the ``missing'' information has 
gone: it has simply vanished into the singularity.  
Of course, one might hope that quantum gravity would eliminate the
singularity, but that need not guarantee that the lost correlations 
would have to reappear for a future observer.

Restricted to black holes, such a loss of unitarity might seem
innocuous.  But there are presumably quantum amplitudes for any
physical process that include virtual black holes, so the effect,
though perhaps tiny, would be pervasive.  It has been argued that 
such violations of unitarity lead to either violations of causality or 
energy nonconservation at an unacceptable level \cite{Banks}, 
but there are known counterexamples \cite{UnruhWald}, and the 
possibility remains open.

\subsection{No black holes}

We really understand the formation of a black hole only as a 
classical process, in which the collapsing matter is typically in 
a complicated thermal state.  It is possible that matter in a pure
state simply does not collapse to form a black hole.  In the
fuzzball picture of section \ref{fuzza}, for instance, the
microscopic states are not black holes; properties such as
horizons appear only statistically as features of ensembles
\cite{Mathurc}.

One can also write down ``phenomenological'' metrics for
which the black hole singularity is replaced by a nonsingular
de Sitter-like region \cite{FrolovVil,Hayward,Frolov,Bardeen}.
In such models, a true horizon never forms, but there can be a
``quasi-horizon'' that looks much like an event horizon for 
a long period of time.  Recall from section \ref{othera} that
a horizon is not required for Hawking radiation --- it is 
enough to have exponential blue shift and a slowly varying 
effective surface gravity --- so for the most part, the
thermodynamics will not change.  But there is no longer
a singularity at which correlations can be lost, and when
the quasi-horizon eventually disappears, the information
in the interior becomes accessible.

Such models share some similarities with the ``remnant''
scenarios discussed below.  In particular, there are unsettled
issues concerning how fast the information can ``escape'' 
after the disappearance of a quasi-horizon.  Nevertheless,
they capture a feature that must be shared by almost any
model of unitary evolution: if the black hole eventually
disappears without any loss of information, then almost by
definition there cannot have been a true event horizon.

\subsection{Remnants and baby universes}

Calculations of Hawking radiation seem likely to be reliable
down to the length scales at which quantum gravitational effects 
become important.  Beyond that scale, though, we do not know 
what to expect.  In particular, it is possible that black hole 
evaporation might stop \cite{Aharanov}, leaving a black hole 
``remnant.'' 
  
If remnants exist, they offer a new resolution to the
information loss problem.  A remnant could be
correlated with the  Hawking radiation, allowing
the combined state to remain pure.  Of course, this requires
that remnants have a huge entropy, that is, a huge number of 
possible states.  This is generally understood to be incompatible
with the AdS/CFT correspondence, and there is a danger of
remnants being overproduced by pair production in the vacuum
\cite{Giddingsx}.  On the other hand, particular exact models 
of black hole evaporation \emph{do} avoid the information
loss problem through a remnant mechanism \cite{Almheiri},
providing at least an existence proof.

Remnants need not have infinite lifetimes; they, too, may
eventually decay into something like Hawking radiation.  There
are, however,  constraints on such a process, if
unitarity is to be preserved \cite{Aharanov,Carlitz}.  As an
estimate of the lifetime \cite{Giddingsy}, suppose the remnant 
has mass $m_R$, and decays by emitting $N$ photons.  Each 
photon carries roughly one bit of information, and if the
process is to preserve unitarity, their total entropy must be
on the order of that of the original black hole.  We
thus require $N\sim GM^2/\hbar$, so each photon should have
an energy and wavelength
\begin{align*}
E \sim \hbar m_R/GM^2, \qquad \lambda \sim GM^2/m_R  .
\end{align*}
For the photons to be uncorrelated, they should be emitted far enough
apart that their wave packets do not significantly overlap.  The total 
time for emission is thus
\begin{align}
\tau \sim N\lambda \sim \left(\frac{M}{m_{Pl}}\right)^4\frac{\hbar}{m_R}  ,
\label{Carlipj2}
\end{align}
typically far longer than the age of the Universe.
By carefully analyzing the entanglement entropy at future null
infinity, Bianchi has found a similar time scale, although with a
bit more flexibility depending on the details of the evaporation
\cite{Bianchix}.

A related approach to the problem is the ``baby 
universe'' scenario \cite{Hawking_baby}.  While we might expect
quantum gravity to eliminate the singularity in figure \ref{Carlipfig5},
the result could plausibly be a new region of the Universe
that is causally disconnected from $\mathcal{I}^+$ --- something
one could perhaps think of as a peculiar, large remnant.  In this case,
unitarity is technically maintained, but it cannot be tested by a single
observer.   

\subsection{Hawking radiation as a pure state}

The most widely held expectation among physicists in this
field is that the information loss problem will be resolved by a
demonstration that Hawking radiation has subtle hidden correlations 
and is actually in a pure state.  This belief gains support from the
AdS/CFT correspondence: in the setting in which we think we best
understand quantum gravity, the boundary conformal field theory
is unitary, and there does not seem to be room in its Hilbert space
for remnants.  At first sight, this idea does not seem unreasonable; 
the density matrix of a pure state can be very close to a thermal 
density matrix \cite{Papadodimas}, differing only by terms of
order $e^{-S}$.  

But while the difference between a pure state and a thermal state
may be very small, it can still be large enough to cause trouble.
\begin{figure}
\centerline{\includegraphics[height=1.3in]{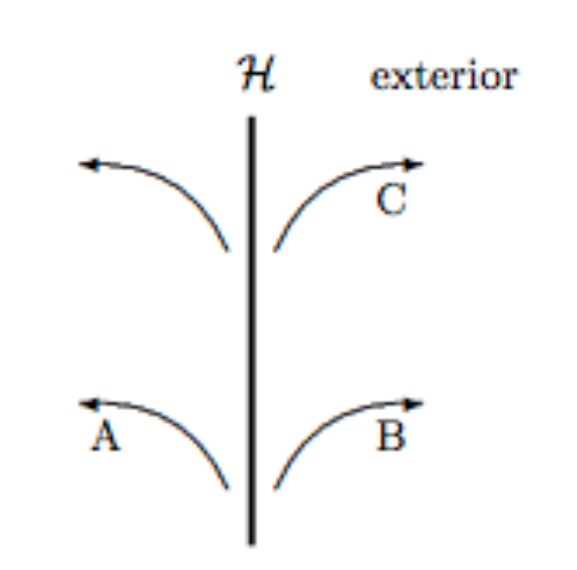}}
\caption{Early and late Hawking quanta emitted from near the
horizon $\mathcal{H}$.}
\label{Carlipfig6}
\end{figure}
Consider two Hawking quanta, one ``early'' and one ``late,'' 
emitted by a black hole, as shown in figure \ref{Carlipfig6}.
As in section \ref{Tempa}, each escaping particle 
is accompanied by a ``partner'' inside the horizon.  In order 
for a freely falling observer to see the standard vacuum near
 $\mathcal{H}$, particle $B$ must initially be 
entangled with its ``partner'' $A$.  This correlation is part of 
the basic structure of the quantum vacuum; it can be violated 
only at the cost of introducing large vacuum expectation values 
of such quantities as the stress-energy tensor.

On the other hand, if the Hawking radiation is to be in a pure 
state in the far future, particle $A$ must be entangled with
other particles in the radiation --- for simplicity, say particle
$C$.  But this violates a fundamental result of quantum mechanics,
``monogamy of entanglement,'' which states that a state
cannot be simultaneously maximally entangled with two
others \cite{monogamy}.  More technically, the problem can be
restated as a violation of the strong subadditivity of entropy
\cite{AMPS}.

Of course, two quantum states that are initially entangled need
not remain so, and it might be possible to transfer the entanglement
of particle $B$ from $A$ to $C$.  But $A$ and $C$ are not causally 
connected --- they are separated by a horizon --- so such a process 
would have to be nonlocal
\cite{Giddingsz}.   One might respond that quantum gravity is
always nonlocal \cite{Jacobsonx,Papadodimasb}, but it remains 
unclear whether ``enough'' nonlocality can be made compatible 
with local effective field theory without doing violence to 
our notions of the scales at which quantum gravitational effects
ought to be important.

For a slightly different perspective \cite{STU}, consider the ``membrane''
description of effective degrees of freedom of section \ref{effecta},
or more generally a holographic picture in which the black hole is 
described by horizon degrees of freedom.  As particle $A$ enters
the black hole, its state should be captured by these horizon 
degrees of freedom, which could then be correlated with the 
later emission of particle $C$.  An infalling observer, on the other
hand, should be able to see particle $A$ inside the horizon.  This
would seem to violate the ``no cloning'' theorem of quantum 
theory \cite{clone}, which prohibits the duplication of a state.
The notion of ``black hole complementarity'' is that this need 
not be a problem, because no single observer can 
measure both states.  But in a much-cited paper, Almheiri {\it et al.}\ 
have disputed this \cite{AMPS}, arguing instead that the entanglement 
across the horizon is likely to be broken; see also Braunstein 
\cite{Braun} and Mathur \cite{Mathur_info} for similar arguments.  The
resulting large expectation values of the stress-energy tensor would
then form a ``firewall'' for any infalling observer.  As Almheiri 
{\it et al.}\ point out, this implies a breakdown of the principle 
of equivalence near a horizon, a rather dramatic claim.  At this 
writing, the controversy is far from being settled.

\section{Conclusion}

Classically, black holes are very nearly the simplest structures in
general relativity.  With the advent of black hole thermodynamics,
however, we have come to see them as highly complex thermal
systems, rarely at equilibrium, with a truly remarkable number of
internal states.  

Perhaps most surprisingly, we have learned that ordinary methods
of general relativity and quantum theory --- quantum field theory
in curved spaces, WKB approximations, semiclassical path integrals,
and the like --- allow us to probe properties of quantum gravitational
states.  Such a claim should be greeted with skepticism, of course.
But over the past few decades we have gathered such a weight of
self-consistent results, based on enough different and independent
approximations, that it seems almost certain we are seeing something 
real.

At the same time, black hole thermodynamics is having a profound
impact on the rest of physics.  The holographic conjecture suggests 
that our fundamental notions of local physics, and perhaps of space
and time, are only low energy approximations.  Attempts to
understand black hole universality, while less sweeping, also point
to the key role of ``boundary'' states.  The information loss problem
has led us to question basic aspects of quantum mechanics, and has
exhibited remarkable connections among very different aspects of
physics.  And perhaps in the future, black hole thermodynamics
will tell us something profound about quantum gravity.
We have interesting times ahead.

\section{Acknowledgments}
This work was supported in part by U.S.\ Department of Energy grant
DE--FG02-91ER40674.  

\begin{appendix}
\section{Classical black holes}

The well-known Schwarzschild and Kerr solutions are
the archetypes of black holes.  For black hole thermodynamics,
though, we need more general configurations: ``dirty'' black holes
whose horizons are distorted by nearby matter, dynamical black holes
formed from collapsing matter, black holes in other spacetime dimensions.
For this, we need a generalization of the notion of a horizon.

A black hole is defined as a ``region of no return,'' an area of
spacetime from which light can never escape.   This idea
can be made more precise---see Hawking and Ellis \cite{HawkingEllis}
for details---by considering a spacetime whose Carter-Penrose 
diagram looks roughly like that of  figure \ref{Carlipfig0a}, in the sense 
that it contains a future null infinity $\mathcal{I}^+$ that describes 
the ``end points at infinity'' of light rays.  Points in the causal past of 
$\mathcal{I}^+$ are ordinary events from which light can escape 
to infinity.  Points that do not lie in the past of $\mathcal{I}^+$ 
are cut off from infinity, and form a black hole region.  The dividing 
line---the boundary of the past of $\mathcal{I}^+$---is the event 
horizon; in figure \ref{Carlipfig0a}, it is the dashed line labeled 
$\mathcal{H}$.

The event horizon defined in this way has many interesting
global properties: for instance, it cannot bifurcate or 
decrease in area  \cite{HawkingEllis}.  But it might be argued that this 
definition does not quite capture the right physics.  The problem can be
traced to the word ``never'' in the phrase ``light can never escape''---%
to truly make such a statement, one must know about the entire future.  
The event horizon is teleological in nature:  
it is determined by ``final causes,'' events in the indefinite future.  

Imagine, for instance, that the Earth is sitting at the center of a highly 
energetic incoming spherical shell of light, one so energetic that 
it has a Schwarzschild radius of one light year.  To be sure, this is 
not likely, but we also cannot rule out the possibility observationally:
no signal can propagate inward faster than such a shell, so we 
would not know of its existence until it reached us.  Suppose now
that the shell is presently two light years away, collapsing inward
toward us at the speed of light.  Step outside and point a flashlight 
up into the sky.  One year from now, the light from your flashlight
will have traveled one light year, where it will meet the collapsing 
shell just as the shell reaches its Schwarzschild radius.  At that 
point, the light will be trapped at the horizon of the newly formed
Schwarzschild black hole, and will be unable to travel any farther 
outward.  In other words, in this scenario we are \emph{already} at 
the event horizon of a black hole, even though we will see no
evidence for this fact until we are suddenly vaporized two years 
from now.

This seem odd; surely Hawking radiation ``now'' should not need to
know about the infinite future.  A number of attempts have been made 
to find more local versions of the event horizon \cite{Boothb},
leading to a variety of more or less useful definitions.  The 
most useful for our purposes is the ``isolated horizon'' \cite{Ashisol},
a locally defined surface with properties suitable for 
describing equilibrium black holes.  

An isolated horizon is a null surface---a surface traced out by light 
rays---whose area remains constant in time, as the horizon of a 
stationary black hole does.  A thought experiment may again be helpful.  
Picture a spherical lattice studded with flashbulbs, set to all flash 
simultaneously in the lattice rest frame.   The bulbs will produce two 
spherical shells of light, one traveling inward and one traveling outward.  
As we know from our experience in nearly flat spacetime, the area of
the ingoing shell will normally decrease with time, while the area 
of the outgoing shell will increase.  If the lattice is placed at the 
horizon of a Schwarzschild black hole, though, it is not hard to 
show that the area of the outgoing shell will remain constant.  
Similarly, if the lattice is placed inside the horizon, \emph{both} 
shells will decrease in area.\footnote{The outgoing shell must
still be outgoing with respect to the lattice, of course; as the lattice 
itself collapses inward, its area will decrease even faster than that 
of the ``outgoing'' shell of light.  This is one way of understanding 
why any object inside a black hole must necessarily collapse inward.}   

To express this idea mathematically, let us first define a 
nonexpanding horizon $\mathcal{H}$ in a $d$-dimensional 
spacetime to be a $(d-1)$-dimensional submanifold such that 
\cite{Ashisol,Ashisolb}
\begin{enumerate}
\item $\mathcal{H}$ is null: that is, its normal vector $\ell_a$
 is a null vector;
\item the expansion of $\mathcal{H}$ vanishes: $\vartheta_{(\ell)}
 = q^{ab}\nabla_a\ell_b = 0$, where $q_{ab}$ is the induced metric
 on $\mathcal{H}$;
\item the field equations hold on $\mathcal{H}$, and $-T^a{}_b\ell^b$ 
is future-directed and causal.
\end{enumerate}
The first condition expresses the fact that the horizon is traced
out by light rays.  The expansion $\vartheta_{(\ell)}$ is
the logarithmic derivative of the area of a cross-section of 
$\mathcal{H}$ \cite{Poisson}, so the second condition implies
time independence of the horizon area.  The third condition 
is a relatively weak prohibition of negative energy at the horizon.

These three conditions imply that
$$\nabla_a\ell^b = \omega_a\ell^b \quad\hbox{on $\mathcal{H}$} $$
for some one-form $\omega_a$.  The surface gravity $\kappa_{(\ell)}$ 
for the normal $\ell^a$---the quantity that appears in the Hawking
temperature (\ref{Carlipa1})---is then defined as
\begin{align}
\kappa_{(\ell)} = \ell^a\omega_a .
\label{CarlipApp1}
\end{align}
The normal $\ell^a$ is not quite unique, though: if $\ell^a$ is a null 
normal to $\mathcal{H}$ and $\lambda$ is an arbitrary function, then
$e^\lambda\ell^a$ is also a null normal to $\mathcal{H}$.  We can 
reduce this ambiguity by demanding further time independence:  a 
weakly isolated horizon is one for which, as an added condition,
we require that
\begin{enumerate}{\setcounter{enumi}{3}}
\item ${\mathcal{L}}_\ell\omega = 0$ on $\cal H$ ,
\end{enumerate}
where $\mathcal{L}$ is the Lie derivative.  
This constraint implies the zeroth law of black hole mechanics, that 
the surface gravity is constant on the horizon.   

Even with this fourth condition, the null normal $\ell^a$ may be 
rescaled by an arbitrary constant.  This also rescales the surface 
gravity, so the numerical value of $\kappa_{(\ell)}$ is not uniquely
determined.  
At first sight this seems to be a terrible feature, but it reflects 
a genuine physical ambiguity: even at a horizon, the choice
of time coordinate is not completely fixed.  In fact, the first law of 
black hole mechanics (\ref{Carlipb1}) \emph{requires} such an 
ambiguity: the scale of the mass $M$ is also fixed only after 
one has normalized the scale of time at infinity.

To clarify this issue, it will help to take a small detour.  For a 
\emph{stationary} black hole, another type of a horizon can be 
defined.  A Killing horizon is a $(d-1)$-dimensional submanifold 
$\mathcal{H}_K$ such that 
\begin{enumerate}
\item some Killing vector $\chi^a$ is null, that is, $\chi_a\chi^a=0$;
and
\item $\mathcal{H}_K$ is itself a null surface, that is, its normal
vector is null.  
\end{enumerate}
If both conditions hold, it follows that $\chi^a$ is itself a null 
normal to $\mathcal{H}_K$, and that
\begin{align}
\chi^a\nabla_a\chi^b = \kappa_{(\chi)}\chi^b 
\quad\hbox{on $\mathcal{H}_K$} .
\end{align}
Note also that the expansion $q^{ab}\nabla_a\chi_b$ is automatically
zero on $\mathcal{H}_K$, simply by virtue of the Killing equation
$\nabla_a\chi_b + \nabla_b\chi_a=0$.

In this case, the null vector $\ell^a$ that defines an isolated horizon
may be chosen to equal $\chi^a$ at $\mathcal{H}_K$, and the
two horizons coincide.  This does not quite solve the problem
of normalization: a Killing vector also requires normalization,
since if $\chi^a$ is a Killing vector, so is $c\chi^a$.  But in the
stationary case, the normalization of $\chi^a$ can be fixed at
infinity, for example by requiring that $\chi^t\sim1$.  In other
words, for stationary black holes one can use global properties 
of the spacetime to adjust clocks at the horizon by comparing 
them with clocks at infinity.  

If, on the other hand, one is only concerned with physics at or 
near the horizon, the normalization of $\kappa_{(\ell)}$ becomes 
more problematic.  One can use known properties of exact solutions 
to express the surface gravity in terms of other quantities at the 
horizon \cite{AshKrish}, thereby fixing $\ell^a$, but so far the 
procedure seems rather artificial.  Alternatively, one can simply
accept the ambiguity, and note that other quantities such as
quasilocal masses defined near $\mathcal{H}$ require a
similar choice of normalization.

As noted in section \ref{prehist}, weakly isolated horizons obey 
the four laws of black hole mechanics, the second law in the  
equilibrium form that the horizon area remains constant.  Generalizations
to dynamical, evolving horizons are also possible \cite{AshKrish}, 
and may be used to prove the inequality version (\ref{Carlipb2})
of the second law.  These generalizations also
provide a potential setting for nonequilibrium black hole 
thermodynamics, allowing us, for instance, to describe  
flows of energy and angular momentum that include the 
contribution of the gravitational field.

\end{appendix}

%\section{Bibliography}\label{secbib}\index{bibliography}


\begin{thebibliography}{999}\addtolength{\itemsep}{-.1ex}
\bibitem{Bekb} J.~D.\ Bekenstein, Black holes and the second law,
\emph{Lett.\ Nuovo Cim.} {\bf 4}, (1972), 737--740.

\bibitem{Bekenstein} J.~D.\ Bekenstein, Black holes and entropy,
\emph{Phys.\ Rev.\ D} {\bf 7},   (1973), 2333--2346.

\bibitem{Hawking} S.~W.\ Hawking, Black hole explosions,
\emph{Nature} {\bf 248}, (1974), 30--31

\bibitem{Hawkingc}  S.~W.\ Hawking, Particle creation by black holes,
\emph{Commun.\ Math.\ Phys.} {\bf 43}, (1975), 199--220

\bibitem{AMPS} A.\ Almheiri, D.\ Marolf, J.\ Polchinski, and J.\ Sully,
Black holes: complementarity or firewalls? \emph{JHEP} {\bf 1302},
(2013), 062, arXiv:1207.3123.

\bibitem{Hawkingb} S.~W.\ Hawking,  Black holes in general relativity,
\emph{Commun.\ Math.\ Phys.} {\bf 25}, (1972), 152--166.

\bibitem{BCH} J.\ M.\ Bardeen, B.\ Carter, and S.\ W.\ Hawking, 
 Commun.\ Math.\ Phys.\ 31, 161 (1973).

\bibitem{Wall} A.~C.\ Wall, Ten proofs of the generalized second 
law, {\emph JHEP} {\bf 0906}, (2009), 021, arXiv:0901.3865.

\bibitem{Ashisol} A.\ Ashtekar, C.\ Beetle, and S.\ Fairhurst,
Isolated horizons: a generalization of black hole mechanics,
\emph{Class.\ Quantum Grav.} {\bf 16}, (1999), L1-L7,
arXiv:gr-qc/9812065.

\bibitem{Wald_Noether} R.~M.\ Wald, Black hole entropy is the 
Noether charge, \emph{Phys.\ Rev. D} {\bf 48}, (1993), 
3427--3431, arXiv:gr-qc/9307038.

\bibitem{Wheeler} J.~A.\ Wheeler and K.\ Ford, \emph{Geons,
Black Holes, and Quantum Foam} (W.~W.~Norton \& Company,
New York, 1998), chap.\ 14.

\bibitem{Waldb}  R.~M.\ Wald, The thermodynamics of black holes,
\emph{Living Rev\ .Rel.} {\bf 4}, (2001), 6, arXiv:gr-qc/9912119.

\bibitem{Wallx}  A.~C.\ Wall, A proof of the generalized second 
law for rapidly changing fields and arbitrary horizon slices,
emph{Phys.\ Rev.\ D} {\bf 85}, (2012), 104049,
arXiv:1105.3445; Erratum-ibid.\ D {\bf 87}, (2013), 069904.

\bibitem{Zeldovich} Ya.~B.\ Zel'dovich and A.~A.\ Starobinsky, 
Particle production and vacuum polarization in an anisotropic 
gravitational field, \emph{Zh.\ Eksp.\ Teor.\ Fiz.} {\bf 61}, 
(1971), 2161--2175; \emph{Sov.\ Phys.\ JETP} {\bf 34}, 
(1972), 1159--1166.

\bibitem{Pretorius} F.\ Pretorius, D.\ Vollick, and W.\ Israel,
An operational approach to black hole entropy, \emph{Phys.\
Rev.\ D} {\bf 57}, (1998), 6311--6316, arXiv:gr-qc/9712085.

\bibitem{Parker} L.\ Parker, Quantized fields and particle 
creation in expanding universes, \emph{Phys.\ Rev.} {\bf 183}, 
(1969), 1057--1068.

\bibitem{Parikh} M.~K.\ Parikh and F.\ Wilczek, Hawking 
radiation as tunneling, \emph{Phys.\ Rev.\ Lett.} {\bf 85}, 
(2000), 5042--5045, arXiv:hep-th/9907001.

\bibitem{BirrellDavies}  N.~D.\ Birrell and P.~C.~W.\ Davies,
\emph{Quantum fields in curved space} (Cambridge University
Press, Cambridge, 1982).

\bibitem{Bogoliubov} N.~N.\ Bogoliubov, 
\emph{Sov.\ Phys.\ JETP} {\bf 7}, (1958), 51.

\bibitem{Traschen} J.~H.\ Traschen, An Introduction to black hole 
evaporation, in eds.\ A.~A.\ Bytsenko and F.~L.\ Williams, 
\emph{Mathematical methods in physics, Proceedings of the 1999 
Londrona Winter School}  (World Scientific, Singapore, 2000), 
arXiv:gr-qc/0010055.

\bibitem{BLSV} C.\ Barcelo, S.\ Liberati, S.\ Sonego, and M.\ Visser,
Minimal conditions for the existence of a Hawking-like flux,
\emph{Phys.\ Rev.\ D} {\bf 83}, (2011), 041501, arXiv:1011.5593.

\bibitem{Page} D.~N.\ Page, Particle emission rates from a 
black hole: massless articles from an uncharged, nonrotating 
hole, \emph{Phys.\ Rev.\ D} {\bf 13}, (1976), 198--206.

\bibitem{Page2} D.~N.\ Page, Hawking radiation and black hole 
thermodynamics, \emph{New J.\ Phys.} {\bf 7}, (2005),
203, arXiv:hep-th/0409024.

\bibitem{Carlip} S.\ Carlip, Black hole thermodynamics and 
statistical mechanics, \emph{Lect.\ Notes Phys.} {\bf 769},
 (2009), 89--123,  arXiv:0807.4520.

\bibitem{Kiefer} C.\ Kiefer, Thermodynamics of black holes and 
Hawking radiation, in eds.\ P.\ Fre, V.\ Gorini, G.\ Magli, and 
U.\ Moschella, \emph{Classical and quantum black holes}
(IOP Publishing, Bristol, 1999).

\bibitem{Schutz} B.\ Schutz, \emph{A first course in general 
relativity}, 2nd edition (Cambridge University Press, Cambridge,
2009), chap.\ 11.5.

\bibitem{Unruhb} W.~G.\ Unruh, Experimental black hole evaporation,
\emph{Phys.\ Rev.\ Lett.} {\bf 46}, (1981), 1351--1353.

\bibitem{Jacobsonb} T.\ Jacobson, Black hole evaporation and 
ultrashort distances, \emph{Phys.\ Rev.\ D} {\bf 44}, (1991),
1731--1739.

\bibitem{Helfer} A.~D.\ Helfer, Do black holes radiate?,
\emph{Rept.\ Prog.\ Phys.} {\bf 66}, (2003), 943--1008,
arXiv:gr-qc/0304042.

\bibitem{Corley}  S.\ Corley, Computing the spectrum of black 
hole radiation in the presence of high frequency dispersion: an 
analytical approach, \emph{Phys.\ Rev.\ D} {\bf 57}, (1998), 
6280--6291, arXiv:hep-th/9710075.

\bibitem{Jacobsonc}  T.\ Jacobson, On the origin of the outgoing 
black hole modes, \emph{Phys.\ Rev.\ D} {\bf 53},  (1996), 
7082--7088, arXiv:hep-th/9601064.

\bibitem{Schutzhold}  W.~G.\ Unruh and R.\ Schutzhold, 
On the universality of the Hawking effect, \emph{Phys.\ Rev.\ D}
{\bf 71},  (2005), 024028, arXiv:gr-qc/0408009.

\bibitem{BLV}  C.\ Barcelo, S.\ Liberati, and M.\ Visser,
Analogue gravity, \emph{Living Rev.\ Rel.} {\bf 14}, (2011), 3,
arXiv:gr-qc/0505065.

\bibitem{Unruhc} W.~G.\ Unruh, Notes on black hole evaporation,
\emph{Phys.\ Rev. D} {\bf 14}, (1976), 870--892.

\bibitem{Visser} M.\ Visser, Hawking radiation without black hole 
entropy, \emph{Phys.\ Rev.\ Lett.} {\bf 80}, (1998), 3436--3439,
arXiv:gr-qc/9712016.

\bibitem{Visser_essential} M.\ Visser, Essential and inessential 
features of Hawking radiation, \emph{Int.\ J.\ Mod.\ Phys.\ D}
{\bf 12}, (2003), 649--661, arXiv:hep-th/0106111.

\bibitem{HartleHawking} J.~B.\ Hartle and S.~W.\ Hawking,
Path integral derivation of black hole radiance, \emph{Phys.\
Rev.\ D} {\bf 13}, (1976), 2188--2203.

\bibitem{Wald} R.~M.\ Wald, On particle creation by black holes,
\emph{Commun.\ Math.\ Phys.}. {\bf 45}, 9--34. (1975).

\bibitem{Parkerb}  L.\ Parker, Probability distribution of particles 
created by a black hole, \emph{Phys.\ Rev.\ D} {\bf 12}, (1975),
1519--1525.

\bibitem{Bisognano} J.~ J.\ Bisognano and E.~H.\ Wichmann, 
On the duality condition for quantum fields, \emph{J.\ Math.\ Phys.}
{\bf 17}, (1976), 303--321.

\bibitem{DeWitt} B.~S.\ DeWitt, Quantum gravity: the new synthesis,
in eds.\ S.~W.\ Hawking and W.\ Israel, \emph{General relativity: an
Einstein centenary survey}, (Cambridge University Press, Cambridge,
1979).

\bibitem{Yu} H.\ Yu and W.\ Zhou, Do static atoms outside a 
Schwarzschild black hole spontaneously excite?, \emph{Phys.\ Rev.\ D}
{\bf 76}, (2007), 044023, arXiv:0707.2613.

\bibitem{Damour} T.\ Damour and R.\ Ruffini, Black hole evaporation 
in the Klein-Sauter-Heisenberg-Euler formalism, \emph{Phys.\ Rev.\ D}
{\bf 14}, (1976), 332--334.

\bibitem{SriPad} K.\ Srinivasan and T.\ Padmanabhan, Particle 
production and complex path analysis, \emph{Phys.Rev. D} {\bf 60}, 
(1999), 024007, arXiv:gr-qc/9812028.

\bibitem{Vanzoreview} L.\ Vanzo, G.\ Acquaviva, and R.\ Di Criscienzo,
Tunneling methods and Hawking's radiation: achievements and prospects,
\emph{Class.\ Quantum Grav.} {\bf 28}, (2011), 183001,
arXiv:1106.4153.

\bibitem{HJ} R.\ Kerner and R.~B.\ Mann, Tunneling, temperature, and 
Taub-NUT black holes, \emph{Phys.\ Rev.\ D} {\bf 73}, (2006),
104010, arXiv:gr-qc/0603019.

\bibitem{Daviesx} P.~C.~W./ Davies, S.~A./ Fulling, and W.~G./ Unruh,
Energy-momentum tensor near an evaporating black hole, \emph{Phys.\
Rev.\ D} {\bf 13}, (1976), 2720--2723.

\bibitem{FMS} P.\ Di Francesco, P.\ Mathieu, and D.\ S{\'e}n{\'e}chal,
 \emph{Conformal field theory} (Springer, New York, 1997).

\bibitem{Christensen} S.~M.\ Christensen and S.~A.\ Fulling,
Trace Anomalies and the Hawking Effect, \emph{Phys.\ Rev.\ D}
{\bf 15}, (1977), 2088--2104.

\bibitem{Robinson} S.~P.\ Robinson and F.\ Wilczek, A relationship 
between Hawking radiation and gravitational anomalies, \emph{Phys.\ 
Rev.\ Lett.} {\bf 95}, (2005), 011303, arXiv:gr-qc/0502074.

\bibitem{Iso} S.\ Iso, H.\ Umetsu, and F.\ Wilczek, Hawking radiation 
from charged black holes via gauge and gravitational anomalies,
\emph{Phys.\ Rev.\ Lett.} {\bf 96}, (2006), 151302, arXiv:hep-th/0602146.

\bibitem{Umetsu} S.\ Iso, H.\ Umetsu, and F.\ Wilczek, Anomalies, 
Hawking radiations and regularity in rotating black holes,
\emph{Phys.\ Rev.\ D} {\bf 74}, (2006), 044017, arXiv:hep-th/0606018.

\bibitem{IMU} S.\ Iso, T.\ Morita, and H.\ Umetsu, Hawking radiation 
via higher-spin gauge anomalies, \emph{Phys.\ Rev.\ D} {\bf 77}, 
(2008), 045007, arXiv:0710.0456. 

\bibitem{BCa} L.\ Bonora, M.\ Cvitan, S.\ Pallua, and I.\ Smoli{\'c},
Hawking fluxes, $W_\infty$ algebra and anomalies, \emph{JHEP} 
{\bf 0812}, (2008), 021, arXiv:0808.2360.

\bibitem{BCb} L.\ Bonora, M.\ Cvitan, S.\ Pallua, and I.\ Smoli{\'c},
Hawking fluxes, fermionic currents, $W_{1+\infty}$ algebra and 
anomalies, \emph{Phys\ .Rev.\ D} {\bf 80}, (2009), 084034,
arXiv:0907.3722.

\bibitem{Kubo} R.\ Kubo, Statistical mechanical theory of irreversible 
processes, \emph{J.\ Phys.\ Soc.\ Japan} {\bf 12}, (1957), 570--586.

\bibitem{Martin} P.~C.\ Martin and J.~S.\ Schwinger, Theory of 
many particle systems, \emph{Phys.\ Rev.} {\bf 115}, (1959),
1342-1373.

\bibitem{Haag} R.\ Haag,  N.~M.\ Hugenholtz, and M.\ Winnink,
On the equilibrium states in quantum statistical mechanics,
\emph{Commun.\ Math.\ Phys.} {\bf 5}, (1967), 215--236.

\bibitem{Gibb}  G.~W.\ Gibbons and M~ J.\ Perry, Black holes in 
thermal equilibrium, \emph{Phys.\ Rev.\ Lett.} {\bf 36}, (1976),
985--987.

\bibitem{Gibc} G.~ W.\ Gibbons and M.~ J.\ Perry, Black holes 
and thermal Green's functions, \emph{Proc.\ Roy.\ Soc.\
Lond.\ A} {\bf 358}, (1978), 467--494.

\bibitem{Feynman}  R.~P.\ Feynman and A.~R.\ Hibbs, emph{Quantum
mechanics and path integrals} (McGraw-Hill, New York, 1965).

\bibitem{Burgess} C.~P.\ Burgess, Quantum gravity in everyday life: 
general relativity as an effective field theory, \emph{Living Rev.\ Rel.}
{\bf 7}, (2004), 5--56, arXiv:gr-qc/0311082.

\bibitem{GibHawk} G.~W.\ Gibbons and S.~W,\ Hawking, Action 
integrals and partition functions in quantum gravity,
\emph{Phys.\ Rev.\ D} {\bf 15}, (1977), 2752--2756.

\bibitem{Regge} C.\ Regge and C.\ Teitelboim, Role of surface 
integrals in the Hamiltonian formulation of general relativity,
\emph{Annals Phys.} {\bf 88}, (1974), 286--318.

\bibitem{BTZa} M.\ Ba{\~n}ados, C.\ Teitelboim, and J.\ Zanelli, 
Black hole entropy and the dimensional continuation of the 
Gauss-Bonnet theorem, \emph{Phys.\ Rev.\ Lett.} {\bf 72},  
(1994), 957--960, arXiv:gr-qc/9309026.

\bibitem{Teitelboim} C.\ Teitelboim, Action and entropy of extreme 
and nonextreme black holes, \emph{Phys.\ Rev.\ D} {\bf 51}, (1995),  
4315--4318, arXiv:hep-th/9410103.

\bibitem{HawkingHorowitz} S.~W.\ Hawking and G.~T.\ Horowitz,
The gravitational Hamiltonian, action, entropy and surface terms,
\emph{Class.\ Quantum Grav.} {\bf 13}, (1996), 1487--1498,
arXiv:gr-qc/9501014.

\bibitem{HawkingHunter} S.~W.\ Hawking and C.\ J.\ Hunter, 
Gravitational entropy and global structure, \emph{Phys.\ Rev.\ D}
{\bf 59}, (1999), 044025, arXiv:hep-th/9808085.

\bibitem{Neiman} Y.\ Neiman, The imaginary part of the gravity 
action and black hole entropy, \emph{JHEP} {\bf 1304}, (2013), 
071, arXiv:1301.7041.

\bibitem{CarTeit} S.\ Carlip and C.\ Teitelboim, The off-shell black 
hole, \emph{Class.\ Quant.\ Grav.} {\bf 12}, (1995), 1699--1704,
arXiv:gr-qc/9312002.

\bibitem{Heisenberg} W.\ Heisenberg and H.\ Euler, Consequences 
of Dirac's theory of positrons, \emph{Z.\ Phys.} {\bf 98}, (1936), 
714--732.

\bibitem{Schwinger} J.~S.\ Schwinger, On gauge invariance and 
vacuum polarization, \emph{Phys.\ Rev.} {\bf 82}, (1951), 664--679.

\bibitem{Garfinkle} D.\ Garfinkle, S.\ B.\ Giddings, and A.\ Strominger,
Entropy in black hole pair production, \emph{Phys.\ Rev.\ D} {\bf 49}, 
(1994), 958--965, arXiv:gr-qc/9306023.

\bibitem{Brown} J.~D.\ Brown, Black hole pair creation and the entropy 
factor, \emph{Phys.\ Rev.\ D} {\bf 51}, (1995), 5725--5731, 
arXiv:gr-qc/9412018.

\bibitem{Dowker} F.\ Dowker, J.~P.\ Gauntlett, D.~A.\ Kastor, and
J.~H.\ Traschen, Pair creation of dilaton black holes, \emph{Phys.\
Rev.\ D} {\bf 49}, (1994), 2909--2917, arXiv:hep-th/9309075.

\bibitem{MannRoss} R.~B.\ Mann and S.~F.\ Ross, Cosmological 
production of charged black hole pairs, \emph{Phys.\ Rev.\ D}
{\bf 52}, (1995), 2254--2265, arXiv:gr-qc/9504015.

\bibitem{Waldbk} R.~M.\ Wald, \emph{Quantum field theory in
curved spacetime and black hole thermodynamics} (University of
Chicago Press, Chicago, 1994).

\bibitem{Kay} B.\ S.\ Kay and R.\ M.\ Wald, Theorems on the 
uniqueness and thermal properties of stationary, nonsingular, 
quasifree states on space-times with a bifurcate Killing horizon,
\emph{Phys.\ Rept.} {\bf 207}, (1991), 41--136.

\bibitem{Israelb} W.\ Israel, Thermo field dynamics of black holes,
\emph{Phys.\ Lett.\ A} {\bf 57}, (1976), 107--110.

\bibitem{Jacobsond} T.\ Jacobson, A note on Hartle-Hawking vacua,
\emph{Phys.\ Rev.\ D} {\bf 50}, (1994), 6031--6032, 
arXiv:gr-qc/9407022.

\bibitem{BTZ} M.\ Banados, C.\ Teitelboim, and J.\ Zanelli, 
The black hole in three-dimensional space-time, \emph{Phys.\ 
Rev.\ Lett.} {\bf 69}, (1992), 1849--1851, arXiv:hep-th/9204099.

\bibitem{BHTZ} M.\ Banados, M.\ Henneaux, C.\ Teitelboim, 
and J.\ Zanelli, Geometry of the (2+1) black hole, \emph{Phys.\
Rev.\ D} {\bf 48}, (1993), 1506--1525, arXiv:gr-qc/9302012.

\bibitem{CarlipBTZ}  S.\ Carlip, The (2+1)-dimensional black hole,
\emph{Class.\ Quantum Grav.} {\bf 12}, (1995), 2853 --2880,
 arXiv:gr-qc/9506079.

\bibitem{Carlipedge} S.\ Carlip, in eds.\ F.\ Khanna and L.\ Vinet,
\emph{Field Theory, Integrable Systems and Symmetries}
(Les Publications CRM, Montreal, 1997), arXiv:gr-qc/9509024.

\bibitem{BrownHen} J.~D.\ Brown and M.\ Henneaux, Central charges 
in the canonical realization of asymptotic symmetries: an example 
from three-dimensional gravity, \emph{Commun.\ Math.\ 
 Phys.} {\bf 104}, 207--226 (1986).

\bibitem{Strominger} A.\ Strominger, Black hole entropy from near 
horizon microstates, \emph{JHEP} {\bf 9802}, 009 (1998),  
arXiv:hep-th/9712251.

\bibitem{BSS} D.\ Birmingham, I.\ Sachs, and S.\ Sen, Entropy of 
three-dimensional black holes in string theory, \emph{Phys.\ Lett.\ B}
 {\bf 424}, 27--2805 (1998), arXiv:hep-th/9801019.

\bibitem{CarlipBTZb} S.\ Carlip, Conformal field theory, 
(2+1)-dimensional gravity, and the BTZ black hole, \emph{Class.\
Quantum Grav.} {\bf 22}, (2005), R85--R124, arXiv:gr-qc/0503022.

\bibitem{ES} R.\ Emparan and I.\ Sachs, Quantization of $\hbox{AdS}_3$ 
black holes in external fields, \emph{Phys.\ Rev.\ Lett.} {\bf 81}, (1998), 
2408--2411, arXiv:hep-th/9806122.

\bibitem{Melnikov} K.\ Melnikov and M.\ Weinstein, On unitary 
evolution of a massless scalar field in a Schwarzschild background: 
Hawking radiation and the information paradox, \emph{Int.\ J.\
Mod.\ Phys.\ D} {\bf 13}, (2004), 1595--1636, arXiv:hep-th/0205223.

\bibitem{Freese} K.\ Freese and C.~T.\ Hill, Covariant functional 
Schr{\"o}dinger formalism and application to the Hawking effect,
\emph{Nucl.\ Phys.\ B} {\bf 255}, (1985), 693--716.

\bibitem{York} J.~W.\ York, Dynamical origin of black hole radiance,
\emph{Phys.\ Rev.\ D} {\bf 28}, (1983), 2929--2945.

\bibitem{Hiscock} W.~A.\ Hiscock and L.~D.\ Weems, Evolution of 
charged evaporating black holes, \emph{Phys.\ Rev.\ D} {\bf 41}, 
(1990), 1142--1151.

\bibitem{Davies} P.~C.~W.\ Davies, The thermodynamic theory of 
black holes, \emph{Proc.\ R. Soc.\ Lond.\ A}. {\bf 353}, 499--521,
(1977).


\bibitem{LyndenBell} D.\ Lynden-Bell and R.\ Wood, The gravo-thermal 
catastrophe in isothermal spheres and the onset of red-giant structure 
for stellar systems, \emph{Mon.\ Not.\ R.\ astr.\ Soc.} {\bf 138},
(1968), 495--525.

\bibitem{Yorkb} J.~W.\ York, Black hole thermodynamics and the
 Euclidean Einstein action, \emph{Phys.\ Rev.\ D} {\bf 33}, (1986), 
2092--2099.

\bibitem{HawkingPage} S.~W.\ Hawking and D.~N.\ Page, Thermodynamics 
of black holes in anti-de Sitter space, \emph{Commun.\ Math.\ Phys.}. 
{\bf 87}, 577--588, (1983).

\bibitem{Witten_Page}  E.\ Witten, Anti-de Sitter space and holography,
\emph{Adv Theor.\ Math.\ Phys.} {\bf 2}, (1998), 253--291,
arXiv:hep-th/9802150,

\bibitem{CarlipVaidya} S.\ Carlip and S.\ Vaidya, Phase transitions 
and critical behavior for charged black holes, \emph{Class.\ Quantum
Grav.} {\bf 20}, (2003). 3827--3838, arXiv:gr-qc/0306054.

\bibitem{HennTeit1} M.\ Henneaux and C.\ Teitelboim, The 
cosmological constant as a canonical variable, \emph{Phys.\ Lett.\ B}
{\bf 143}, (1984), 415--420.

\bibitem{Caldarelli} M.~M.\ Caldarelli, G.\ Cognola, and D.\ Klemm,
thermodynamics of Kerr-Newman-AdS black holes and conformal 
field theories, \emph{Class.\ Quantum Grav.} {\bf 17}. (2000), 
399--420, arXiv:hep-th/9908022.

\bibitem{Sekiwa} Y.\ Sekiwa, Thermodynamics of de Sitter black
 holes: thermal cosmological constant, \emph{Phys.\ Rev.\ D}
{\bf 73}, (2006), 084009, arXiv:hep-th/0602269.

\bibitem{KRT} D.\ Kastor, S.\ Ray, and J.\ Traschen, Enthalpy and 
the mechanics of AdS black holes, \emph{Class.\ Quantum Grav.}
{\bf 26}, (2009), 195011, arXiv:0904.2765.

\bibitem{Cvetic} M.\ Cvetic, G.~W.\ Gibbons, D.\ Kubiznak, and
C.~N.\ Pope, Black hole enthalpy and an entropy Inequality for the 
thermodynamic volume, \emph{Phys.\ Rev.\ D} {\bf 84}, (2011),
024037, arXiv:1012.2888.

\bibitem{Dolan} B.~P.\ Dolan, The cosmological constant and 
the black hole equation of state, \emph{Class.\ Quantum Grav.}
{\bf 28}, (2011), 125020, arXiv:1008.5023.

\bibitem{MannKub}  D.\ Kubiznak and R.~B.\ Mann, Black hole
chemistry, arXiv:1404.2126

\bibitem{DubSib} S.~L.\ Dubovsky and S.~M.\ Sibiryakov, Spontaneous
breaking of Lorentz invariance, black holes and perpetuum mobile of
the 2nd kind, \emph{Phys.\ Lett.\ B}. {\bf 638}, 509--514, (2006).

\bibitem{Eling} C.\ Eling, B.~Z.\ Foster, T.\ Jacobson, and A.~C.\ Wall,
Lorentz violation and perpetual motion, \emph{Phys.\ Rev.\ D}. {\bf 75},
101502(R), (2007).

\bibitem{Barvinsky} A.\ Barvinsky, S.\ Das, and G.\ Kunstatter,
Quantum mechanics of charged black holes, \emph{Phys.\ Lett.\ B}
{\bf 517}, (2001), 415--420, arXiv:hep-th/0102061.

\bibitem{Hod} S.\ Hod, Bohr's correspondence principle and the 
area spectrum of quantum black holes, \emph{Phys.\ Rev.\ Lett.}
{\bf 81}, (1998), 4293--4296, arXiv:gr-qc/9812002.

\bibitem{Siopsis} G.\ Siopsis, Analytic calculation of quasi-normal 
modes, \emph{Lect.\ Notes Phys.} {\bf 769}, (2009), 471--508,
arXiv:0804.2713.

\bibitem{Magg} M.\ Maggiore, The physical interpretation of 
the spectrum of black hole quasinormal modes, \emph{Phys.\ 
Rev.\ Lett.} {\bf 100}, (2008), 141301, arXiv:0711.3145.

\bibitem{Medved} A.~J.~M.\ Medved, On the `universal' quantum 
area spectrum, \emph{Mod.\ Phys.\ Lett.\ A} {\bf 24}, (2009), 
2601--2609, arXiv:0906.2641.

\bibitem{Sorkin} R.~D.\ Sorkin, On the entropy of the vacuum outside 
a horizon, in \emph{Tenth International Conference on General Relativity 
and Gravitation}, vol. II, 734--736 (1983), arXiv:1402.3589.

\bibitem{Bomb} L.\ Bombelli, R.~K.\ Koul, J.\ Lee, and R.~D.\ 
 Sorkin, A quantum source of entropy for black holes, \emph{Phys.\ 
Rev.\ D}. {\bf 34}, 373--383 (1986).

\bibitem{Sred} M.\ Srednicki, Entropy and area, \emph{Phys.\ Rev.\ 
Lett.} {\bf 71}, 666--669 (1993), arXiv:hep-th/9303048.

\bibitem{FroNov} V.\ Frolov and I.\ Novikov, Dynamical origin
of the entropy of a black hole, \emph{Phys.\ Rev.\ D} {\bf 48},
4545--4551 (1993), arXiv:gr-qc/9309001.

\bibitem{tHooft} G.\ 't Hooft, On the quantum structure of a black hole,
\emph{Nucl.\ Phys.} {\bf B256}, 727--745, (1985).

\bibitem{SussUg} L.\ Susskind and J.\ Uglum, Black hole entropy
 in canonical quantum gravity and superstring theory, \emph{Phys.\ Rev.}
{\bf  D50}, 2700--2711, (1994), arXiv:hep-th/9401070.

\bibitem{Coop} J.~H. Cooperman and M.~A. Luty, Renormalization of 
entanglement entropy and the gravitational effective action,
arXiv:1302.1878.

\bibitem{Ryu} S.\ Ryu and T.\ Takayanagi, Holographic derivation of 
entanglement entropy from AdS/CFT, \emph{Phys.\ Rev.\ Lett.} {\bf 96}, 
181602 (2006), arXiv:hep-th/0603001.

\bibitem{Hubeny} V.~E.\ Hubeny, M.\ Rangamani, and T.\ Takayanagi,
A covariant holographic entanglement entropy proposal, \emph{ JHEP}.
{\bf 0707}, 062 (2007), arXiv:0705.0016.

\bibitem{Lewkowycz} A.\ Lewkowycz and J.\ Maldacena, Generalized 
gravitational entropy, \emph{JHEP} {\bf 1308}, (2013), 090,
arXiv:1304.4926.

\bibitem{Emparanb} R.\ Emparan, Black hole entropy as entanglement 
entropy: a holographic derivation, \emph{JHEP}. {\bf 012} 0606 (2006),  
arXiv:hep-th/0603081.

\bibitem{StromVafa} A.\ Strominger and C.\ Vafa, Microscopic origin 
of the Bekenstein-Hawking entropy, \emph{Phys.\ Lett.\ B}. {\bf 379},
99--104, (1996), arXiv:hep-th/9601029.

\bibitem{Peet} A.~W.\ Peet, TASI lectures on black holes in string theory,
in eds.\ J.\ Harvey, S.\ Kachru, and E.\ Silverstein, \emph{TASI 99: Strings, 
branes, and gravity}.  (World  Scientific, Singapore, 2001), 
arXiv:hep-th/0008241.

\bibitem{Das} S.~R.\ Das and S.~D.\ Mathur, The quantum physics of 
black holes: results from string theory, \emph{Ann.\ Rev.\ Nucl.\
 Part.\ Sci.} {\bf 50}, 153--206, (2000), arXiv:gr-qc/0105063.

\bibitem{Mal} J.~Maldacena and A.~Strominger, Black hole greybody 
factors and D-brane spectroscopy, \emph{Phys.\ Rev.} {\bf  D55}, 
861--870, (1997), arXiv:hep-th/9609026.

\bibitem{Mathur}  S.~D.\ Mathur, The fuzzball proposal for black holes: 
an elementary review, \emph{Fortsch.\ Phys.} {\bf 53}, (2005),
793--827, arXiv:hep-th/0502050.

\bibitem{Mathurb} S.~D.\ Mathur, The quantum structure of black holes,
\emph{Class.\ Quantum Grav.} {\bf 23}, (2006), R115--R168,
arXiv:hep-th/0510180.

\bibitem{Mathurc} S.~D.\ Mathur, What exactly is the information 
paradox?, \emph{Lect.\ Notes Phys.} {\bf 769}, (2009), 3--48,
arXiv:0803.2030.

\bibitem{Chowdhury} B.~D.\ Chowdhury and S.~D.\ Mathur,
Radiation from the non-extremal fuzzball, \emph{Class.\ Quantum 
Grav.} {\bf 25}, 135005, (2008), arXiv:0711.4817.

\bibitem{Malda} J.~M.\ Maldacena, The large N limit of superconformal 
field theories and supergravity, \emph{Adv.\ Theor.\ Math.\ Phys.}
{\bf 2},  231--252, (1998), arXiv:hep-th/9711200.

\bibitem{AGMOO} O.\ Aharony, S.~S.\ Gubser, J.~M.\ Maldacena, H.\
Ooguri, and Y.\ Oz, Large N field theories, string theory and gravity,
\emph{Phys.\ Rept.} {\bf 323}, 183--386, (2000), arXiv:hep-th/9905111. 

\bibitem{Skenderisb} K.\ Skenderis, Black holes and branes in string 
theory, \emph{Lect.\ Notes Phys.} {\bf 541}, (2000), 325--364,
arXiv:hep-th/9901050.

\bibitem{Ashisolb} A.\ Ashtekar, C.\ Beetle, and S.\ Fairhurst,
Mechanics of isolated horizons, \emph{Class.\ Quantum Grav.} {\bf 17}, 
(2000), 253--298, arXiv:gr-qc/9907068.

\bibitem{Corichi} A.~Corichi,  J.\ Diaz-Polo, and E.~Fernandez-Borja,
Black hole entropy quantization, \emph{Phys.\ Rev.\ Lett.} {\bf 98},
 (2007), 181301,  arXiv:gr-qc/0609122.

\bibitem{Agullo} I.\ Agullo, J.~Fernando Barbero, E.~F.\ Borja, J.\ Diaz-Polo,
and E.~J.~S.\ Villase{\~n}or, Detailed black hole state counting in loop 
quantum gravity, emph{Phys.\ Rev.\ D} {\bf 82}, (2010), 084029,
arXiv:1101.3660.

\bibitem{Jacobsone} T.\ Jacobson, Renormalization and black hole 
entropy in loop quantum gravity, \emph{Class.\ Quantum Grav.} {\bf 24},
4875--4879,  (2007),  arXiv:0707.4026.

\bibitem{Alexandrov} S.\ Alexandrov and E.\ R.\ Livine, SU(2) loop 
quantum gravity seen from covariant theory, \emph{Phys.\ Rev.\ D}
{\bf 67}, 044009, (2003), arXiv:gr-qc/0209105.

\bibitem{Krasnov}  K.~V.\ Krasnov, Counting surface states in the 
loop quantum gravity, \emph{Phys.\ Rev.\ D} {\bf 55}, (1997),
3505--3513, arXiv:gr-qc/9603025.

\bibitem{Rovellib}  C.\ Rovelli, Black hole entropy from loop 
quantum gravity, \emph{Phys.\ Rev.\ Lett.} {\bf 77}, (1996),
3288--3261, arXiv:gr-qc/9603063.

\bibitem{ABCK} A.\ Ashtekar, J.~C.\ Baez, A.\ Corichi, and K.\ Krasnov,
Quantum geometry and black hole entropy, \emph{Phys.\ Rev.\ Lett.}
{\bf 80}, (1998), 904--907, arXiv:gr-qc/9710007.

\bibitem{ABK} A.\ Ashtekar, J.~C.\ Baez, and K.\ Krasnov, 
Quantum geometry of isolated horizons and black hole entropy,
\emph{Adv.\ Theor.\ Math.\ Phys.} {\bf 4}, (2000), 1--94, 
arXiv:gr-qc/0005126.

\bibitem{Witten_Jones} E.\ Witten, Quantum field theory and the
Jones polynomial, \emph{Commun.\ Math.\ Phys.} {\bf 121}, 
(1989), 351--399.

\bibitem{Domagala} M.\ Domagala and J.\ Lewandowski, Black hole 
entropy from quantum geometry, \emph{Class.\ Quantum Grav.}
{\bf 21}, (2004), 5233--5244, arXiv:gr-qc/0407051.

\bibitem{Meissner} K.~A.\ Meissner, Black hole entropy in loop 
quantum gravity, \emph{Class.\ Quantum Grav.} {\bf 21},
(2004),  5245--5252, arXiv:gr-qc/0407052. 

\bibitem{AshtekarLew} A.\ Ashtekar and J.\ Lewandowski, 
Background independent quantum gravity: A Status report
\emph{Class.\ Quantum Grav.} {\bf 21}, (2004), R53--R152,
arXiv:gr-qc/0404018.

\bibitem{Engle} A.\ Ashtekar, J.\ Engle, and C.\ Van Den Broeck,
Quantum horizons and black hole entropy: Inclusion of distortion 
and rotation, \emph{Class.\ Quantum Grav.} {\bf 22}, (2005),
L27--L34, arXiv:gr-qc/0412003.

\bibitem{Livine} E~R.\ Livine and D.~R.\ Terno, Quantum 
black holes: entropy and entanglement on the horizon
\emph{Nucl.\ Phys.\ B} {\bf 741}, (2006), 131--161,
arXiv:gr-qc/0508085.

\bibitem{Bianchi} E.\ Bianchi, Entropy of non-extremal black holes 
from loop gravity, arXiv:1204.5122.

\bibitem{Frodden} E.\ Frodden, A.\ Ghosh, and A.\ Perez, Quasilocal 
first law for black hole thermodynamics, \emph{Phys.\ Rev.\ D} {\bf 87}, 
121503, (2013), arXiv:1110.4055.

\bibitem{Ghosh}  A.\ Ghosh and A.\ Perez, Black hole entropy and
isolated horizons thermodynamics, \emph{Phys.\ Rev.\ Lett.} {\bf 27},
(2011), 241301, arXiv:1107.1320.

\bibitem{Perez} A.\ Ghosh, K.\ Noui, and A.\ Perez, Statistics, 
holography, and black hole entropy in loop quantum gravity,
arXiv:1309.4563.

\bibitem{Geiller} E.\ Frodden, M.\ Geiller, K.\ Noui, and A.\ Perez,
Black hole entropy from complex Ashtekar variables, arXiv:1212.4060.

\bibitem{Achour} J.~B.\ Achour, A.\ Mouchet, and K.\ Noui,
Analytic continuation of black hole entropy in loop quantum 
gravity, arXiv:1406.6021. 

\bibitem{Carlip_loop} S.~Carlip, Four-dimensional entropy from 
three-dimensional gravity, \emph{ Phys.\ Rev.\ Lett.} {\bf 115}, 
(2015), 071302, arXiv:1503.02981.

\bibitem{Sakharov}  A.\ D.\ Sakharov, Vacuum quantum fluctuations 
in curved space and the theory of gravitation, \emph{Sov.\ Phys.\ Dokl.}
{\bf 12}, (1968), 1040--1041, reprinted in \emph{Gen.\ Rel.\ Grav.} 
{\bf 32}, (2000), 365--367.

\bibitem{Adler} S.\ L.\ Adler, Einstein gravity as a symmetry breaking 
effect in quantum field theory, \emph{Rev.\ Mod.\ Phys.} {\bf 54}, (1982),
729--766.   Erratum ibid.\ {\bf 55}, (1983), 837.

\bibitem{Frolovx} V~ P.\ Frolov, D.~V.\ Fursaev, and A.~I.\ Zelnikov,
Statistical origin of black hole entropy in induced gravity,
\emph{Nucl.\ Phys.\ B} {\bf 486}, (1997), 339--352,
arXiv:hep-th/9607104.

\bibitem{Frolovb} V~ P.\ Frolov and D.~V.\ Fursaev, Mechanism of 
generation of black hole entropy in Sakharov's induced gravity,
\emph{Phys.\ Rev.\ D} {\bf 56}, (1997),  2212--225,
arXiv:hep-th/9703178.

\bibitem{Frolovc} V~ P.\ Frolov, D.\ Fursaev, and A.\ Zelnikov, 
CFT and black hole entropy in induced gravity, \emph{JHEP}
{\bf 0303}, (2003), 038, arXiv:hep-th/0302207.

\bibitem{Sorkin_caus}  L.\ Bombelli, J.\ Lee, D.\ Meyer, and R.\ Sorkin,
Space-time as a causal set, \emph{Phys.\ Rev.\ Lett.} {\bf 59}, (1987), 
521--524.

\bibitem{Rideout} D.\ Rideout and S.\ Zohren, Evidence for an 
entropy bound from fundamentally discrete gravity, \emph{Class.\ 
Quantum Grav.} {\bf 23}, (2006),  6195--6213, arXiv:gr-qc/0606065.

\bibitem{Dou} D.\ Dou and R.~D.\ Sorkin, Black hole entropy as 
causal links, \emph{Found.\ Phys.} {\bf 33}, (2003), 279--296,
arXiv:gr-qc/0302009.

\bibitem{Zurek} W.\ H.\ Zurek and K.\ S.\ Thorne, Statistical 
mechanical origin of the entropy of a rotating, charged black 
hole, \emph{Phys.\ Rev.\ Lett.} {\bf 54}, (1985), 2171--2175.

\bibitem{Carliplog} S.\ Carlip, Logarithmic corrections to black hole entropy 
from the Cardy formula, \emph{Class.\ Quantum Grav.} {\bf 17}, (2000), 
4175--4186, arXiv:gr-qc/0005017.

\bibitem{Ghoshb} A.\ Ghosh and P.\ Mitra, A bound on the log correction to 
the black hole area law, \emph{Phys.\ Rev.\ D} {\bf 71}, (2005), 027502,
arXiv:gr-qc/0401070.

\bibitem{Medvedx} A.~J.~M.\ Medved and E.~C.\ Vagenas, When conceptual 
worlds collide: the GUP and the BH entropy, \emph{Phys.\ Rev.\ D} {\bf 70}, 
(2004), 124021, arXiv:hep-th/0411022.

\bibitem{Sen} A.\ Sen, Logarithmic corrections to Schwarzschild and other 
non-extremal black hole entropy in different dimensions, \emph{JHEP}
{\bf 04}, (2013), 156, arXiv:1205.0971.

\bibitem{ENPP} J.\ Engle, K.\ Noui, A.\ Perez, and D.\ Pranzetti,
The SU(2) black hole entropy revisited, \emph{JHEP} {\bf  1105}, (2011),
016, arXiv:1103.2723.

\bibitem{Gour} G.\ Gour and  A.~J.~M.\ Medved, Thermal fluctuations 
and black hole entropy, \emph{Class.\ Quantum Grav.} {\bf 20}, (2003), 
3307--3326, arXiv:gr-qc/0305018.

\bibitem{Torre} C.~G.\ Torre and I.~M.\ Anderson, Symmetries of 
the Einstein equations, \emph{Phys.\ Rev.\ Lett.} {\bf 70}, (1993), 
3525--3529, arXiv:gr-qc/9302033.

\bibitem{Hooft_holo}  G.\ 't Hooft, Dimensional reduction in 
quantum gravity, in eds.\  A.\ Ali, J.\ Ellis, and S.\ Randjbar-Daemi,
\emph{Salamfestschrift: a collection of talks}  (World Scientific, 
Singapore, 1993), arXiv:gr-qc/9310026.

\bibitem{Suss_holo} L.\ Susskind, The world as a hologram,
\emph{J.\ Math.\ Phys.} {\bf 36}, (1995), 6377--6396,
arXiv:hep-th/9409089.

\bibitem{Yurtsever} U.\ Yurtsever, The holographic entropy 
bound and local quantum field theory, \emph{Phys.\ Rev.\ Lett.} 
{\bf 91}, (2003), 041302, arXiv:gr-qc/0303023.

\bibitem{Hsu} S.~D.~H.\ Hsu and D.\ Reeb, Black hole entropy, 
curved space and monsters, \emph{Phys.\ Lett.\ B} {\bf 658}, 
(2008), 244--248, arXiv:0706.3239. 

\bibitem{Flanagan} E.~E.\ Flanagan, D.\ Marolf, and R.~M.\ Wald,
Proof of classical versions of the Bousso entropy bound and of the 
generalized second law, \emph{Phys.\ Rev.\ D} {\bf 62}, (2000),
084035, arXiv:hep-th/9908070.

\bibitem{Bousso} R.\ Bousso, A covariant entropy conjecture,
\emph{JHEP} {\bf 9907}, (1999), 004, arXiv:hep-th/9905177.

\bibitem{Boussob} R.\ Bousso, The holographic principle, 
\emph{Rev.\ Mod.\ Phys.} {\bf 74}, (2002),  825--874, 
arXiv:hep-th/0203101.

\bibitem{BoussoFlan} R.\ Bousso, E.~E.\ Flanagan, and D.\ Marolf,
Simple sufficient conditions for the generalized covariant 
entropy bound, \emph{Phys.\ Rev.\ D} {\bf 68}, (2003), 064001.
arXiv:hep-th/0305149.

\bibitem{BoussoMal} R.\ Bousso, H.\ Casini, Z.\ Fisher, and 
J.\ Maldacena, Proof of a quantum Bousso bound, arXiv:1404.5635.

\bibitem{Cardy} J.~l.\ Cardy, Operator content of two-dimensional 
conformally invariant theories, \emph{Nucl.\ Phys.\ B} {\bf 270}, 
(1986), 186--204.

\bibitem{Cardy2}  H.~W.~J.\ Bl{\"o}te, J.~A.\ Cardy, and M.~P.\ 
Nightingale, Conformal invariance, the central charge, and 
universal finite size amplitudes at criticality, \emph{Phys.\ Rev.\ 
Lett.} {\bf 56}, (1986), 742--745.

\bibitem{Camblong} H.~E.\ Camblong and C.~R.\ Ord{\'o}{\~n}ez, 
Black hole thermodynamics from near-horizon conformal quantum 
mechanics, \emph{Phys.\ Rev.\ D} {\bf 71}, (2005), 104029,
arXiv:hep-th/0411008.

\bibitem{Carlipcfta} S.\ Carlip, Black hole entropy from conformal 
field theory in any dimension, \emph{Phys.\ Rev.\ Lett.} {\bf 82}, 
(1999), 2828--2831, arXiv:hep-th/9812013.

\bibitem{Carlipcftb} S.\ Carlip, Entropy from conformal field 
theory at Killing horizons, \emph{Class.\ Quantum Grav.} {\bf 16}, 
(1999), 3327--3348, arXiv:gr-qc/9906126.

\bibitem{Carlipcftc} S.\ Carlip, Extremal and nonextremal Kerr/CFT 
correspondences, \emph{JHEP} {\bf 1104}, (2011), 076, 
arXiv:1101.5136;  Erratum-ibid. {\bf 1201} (2012) 008.

\bibitem{Carlipcftd} S.\ Carlip, Effective conformal descriptions
of black hole entropy, \emph{Entropy} {\bf 13}, (2011), 
1355--1379, arXiv:1107.2678. 

\bibitem{GHSS} M.\ Guica, T.\ Hartman, W.\ Song, and A.\ Strominger,
The Kerr/CFT correspondence, \emph{Phys.\ Rev.\ D} {\bf 80}, 
(2009), 124008, arXiv:0809.4266. 

\bibitem{Carlip_wouldbe} S.\ Carlip, Statistical mechanics and 
black hole thermodynamics, \emph{Nucl.\ Phys.\ Proc.\ Suppl.}
{\bf 57}, (1997), 8--12, arXiv:gr-qc/9702017.

\bibitem{Goldstone}  J.\ Goldstone, A.\ Salam, and S.\ Weinberg, 
Broken symmetries, \emph{Phys.\ Rev.} {\bf 127}, (1962), 965--970.

\bibitem{Carlipasymp} S.\ Carlip, Dynamics of asymptotic diffeomorphisms 
in (2+1)-dimensional gravity, \emph{Class.\ Quantum Grav.} {\bf 22},
(2005), 3055--3060, arXiv:gr-qc/0501033.

\bibitem{membrane}  K.~S.\ Thorne, R~ H.\ Price, and D.~A.\ Macdonald,
\emph{Black holes: the membrane paradigm} (Yale University Press, 
New Haven, 1986).

\bibitem{Parikhz} M.\ Parikh and F.\ Wilczek, An action for black hole 
membranes, \emph{Phys.\ Rev.\ D} {\bf 58}, (1998), 064011,
arXiv:gr-qc/9712077.

\bibitem{Hawking_info} S.~W.\ Hawking, Breakdown of predictability 
in gravitational collapse, \emph{Phys.\ Rev.\ D} {\bf 14}, (1976),
2460--2473.

\bibitem{Banks} T.\ Banks, L.\ Susskind, and M.~E.\ Peskin, 
Difficulties for the evolution of pure states Into mixed states,
\emph{Nucl.\ Phys.\ B} {\bf 244}, (1984), 125--134.

\bibitem{UnruhWald} W.~G.\ Unruh and R.~M.\ Wald, On evolution 
laws taking pure states to mixed states in quantum field theory,
\emph{Phys.\ Rev.\ D} {\bf 52},  (1995), 2176--2182, 
arXiv:hep-th/9503024.

\bibitem{FrolovVil} V.~P.\ Frolov and G.~A.\ Vilkovisky,
Spherically symmetric collapse in quantum gravity,
\emph{Phys.\ Lett.| B} {\bf 106}, (1981), 307--313.

\bibitem{Hayward} S.~A.\ Hayward, Formation and evaporation 
of regular black holes, \emph{Phys.\ Rev.\ Lett.} {\bf 96}, (2006), 
031103, arXiv:gr-qc/0506126.

\bibitem{Frolov} V.~P.\ Frolov, Information loss problem and 
a `black hole' model with a closed apparent horizon, 
\emph{JHEP} {\bf 1405}, (2014), 049, arXiv:1402.5446.

\bibitem{Bardeen}  J.~M.\ Bardeen, Black hole evaporation 
without an event horizon,  arXiv:1406.4098.

\bibitem{Aharanov} Y.\ Aharonov, A.\ Casher, and S.\ Nussinov,
The unitarity puzzle and Planck mass stable particles,
\emph{Phys.\ Lett.\ B} {\bf 191}, (1987), 51--55.

\bibitem{Giddingsx} S.~B.\ Giddings, Constraints on black hole 
remnants, \emph{Phys.\ Rev.\ D} {\bf 49}, (1994), 947--957.

\bibitem{Almheiri} A.\ Almheiri and J.\ Sully, An uneventful 
horizon in two dimensions, \emph{JHEP} {\bf 1402}, (2014),
108, arXiv:1307.8149.

\bibitem{Carlitz} R.~D.\ Carlitz and R.~S.\ Willey, The
lifetime of a black hole, \emph{Phys.\ Rev.\ D} {\bf 36},
(1987), 2336--2341.

\bibitem{Giddingsy} S.~B.\ Giddings, Quantum mechanics of 
black holes, in eds.\  E.\ Gava, A.\ Masiero, K.~S.\ Narain, 
S.\ Randjbar-Daemi, and Q. \ Shafi, \emph{Summer school in
high energy physics and cosmology: proceedings} (World
Scientific, Singapore, 1995), arXiv:hep-th/9412138.

\bibitem{Bianchix} E.\ Bianchi, talk at Peyresq Physics 11, 
June 2014.

\bibitem{Papadodimas} K.\ Papadodimas and S.\ Raju,
The unreasonable effectiveness of exponentially suppressed 
corrections in preserving information, \emph{Int.\ J.\ Mod.\ Phys.\ D}
{\bf 22},  (2013,) 1342030.

\bibitem{Hawking_baby} S.~W.\ Hawking, Baby universes, 
\emph{Mod.\ Phys.\ Lett.\ A} {\bf 5}, (1990), 453--466.

\bibitem{monogamy} D.\ Yang, A simple proof of monogamy of
entanglement, \emph{Phys.\ Lett.| A} {\bf 360}, (2006), 249--250,
arXiv:quant-ph/0604168.

\bibitem{Giddingsz} S.~B.\ Giddings, Nonlocality versus complementarity: 
a conservative approach to the information problem, \emph{Class.\
Quantum Grav.} {\bf 28}, (2011), 025002, arXiv:0911.3395.

\bibitem{Jacobsonx} T.\ Jacobson, Boundary unitarity and the black 
hole information paradox, \emph{nt.\ J.\ Mod.\ Phys.\ D} {\bf 22}, 
(2013), 1342002, arXiv:1212.6944.

\bibitem{Papadodimasb}  K.\ Papadodimas and S.\ Raju,
The black hole interior in AdS/CFT and the information paradox,
\emph{Phys.\ Rev.\ Lett.} {\bf 112}, (2014), 051301,
arXiv:1310.6334.

\bibitem{STU} L.\ Susskind, L.\ Thorlacius, and J.\ Uglum,
The stretched horizon and black hole complementarity,
\emph{Phys.\ Rev.\ D} {\bf 48}, (1993), 3743--3761,
arXiv:hep-th/9306069.

\bibitem{clone} W.~K.\ Wootters and W.~H.\ Zurek, A single 
quantum cannot be cloned, \emph{Nature} {\bf 299}, (1982), 
802--803.

\bibitem{Braun}  S.~L.\ Braunstein, S.\ Pirandola, and K.\ Zyczkowski,
Better late than never: Information retrieval from black holes,
\emph{Phys.\ Rev.\ Lett.} {\bf 110},  (2013), 101301,
arXiv:0907.1190.

\bibitem{Mathur_info} S.~D.\ Mathur, The information paradox: 
a pedagogical introduction, \emph{Class.\ Quantum Grav.} {\bf 26},
(2009),  224001, arXiv:0909.1038.

\bibitem{HawkingEllis}  S.~W.\ Hawking and G.~F.~R.\ Ellis,
\emph{The large scale structure of space-time} (Cambridge
University Press, Cambridge, 1973).

\bibitem{Boothb} I.\ Booth, Black hole boundaries, \emph{Can.\
J.\ Phys.} {\bf 83}, (2005), 1073-1099, arXiv:gr-qc/0508107.

\bibitem{Poisson} E.\ Poisson, \emph{A relativist's toolkit} (Cambridge
University Press, Cambridge, 2004).

\bibitem{AshKrish} A.\ Ashtekar and B.\ Krishnan, Isolated and 
dynamical horizons and their applications, \emph{Living Rev.\ Rel.}
{\bf 7}, (2004), 10, arXiv:gr-qc/0407042.

\end{thebibliography}
\end{document}